\documentclass[%
 reprint,
superscriptaddress,
preprintnumbers,  % Please leave this one on
nofootinbib,
 amsmath,amssymb,
 aps,
]{revtex4-1}

\usepackage{graphicx}% Include figure files
\usepackage{dcolumn}% Align table columns on decimal point
\usepackage{bm}% bold math
\usepackage{natbib}
\usepackage{amsmath}
\usepackage{amssymb}
\usepackage{amsfonts}
\usepackage{xfrac}
\usepackage{graphicx}
\usepackage{subfigure}
\usepackage{hyperref}
\usepackage[dvipsnames]{xcolor}
\usepackage[normalem]{ulem}
\newcommand{\diff}{\mathrm{d}}
\newcommand{\defeq}{:=}

\begin{document}

\preprint{TTK-20-44}

\title{Bayesian reconstruction of the inflaton's speed of sound using CMB data}

\author{Guadalupe Ca\~nas-Herrera}
\email{canasherrera@lorentz.leidenuniv.nl}
\affiliation{Lorentz Institute for Theoretical Physics, Leiden University, PO Box 9506, Leiden 2300 RA, The Netherlands}
\affiliation{Leiden Observatory, Leiden University, PO Box 9506, Leiden 2300 RA, The Netherlands}
\author{Jes\'us Torrado}
\email{torrado@physik.rwth-aachen.de}
\affiliation{Institute for Theoretical Particle Physics and Cosmology (TTK), RWTH Aachen University, D-52056 Aachen, Germany}
\affiliation{Dept.\ of Physics \& Astronomy, University of Sussex, Brighton BN1 9QH, UK}
\author{Ana Ach\'ucarro}
\email{achucar@lorentz.leidenuniv.nl}
\affiliation{Lorentz Institute for Theoretical Physics, Leiden University, PO Box 9506, Leiden 2300 RA, The Netherlands}
\affiliation{Dept.\ of Theoretical Physics, University of the Basque Country UPV-EHU, 48080 Bilbao, Spain}
\date{\today}% It is always \today, today,

\begin{abstract}
We update the search for features, due to transient reductions in inflaton's speed of sound, in the Cosmic Microwave Background (CMB) angular power spectrum using Planck 2018 temperature, polarization and lensing data. We develop a new methodology to test more flexible templates to reconstruct the reduction of the speed of sound based on Gaussian Processes. We formally derive a dynamical prior for the shape of the reduction using a \textit{maximum-entropy} approach to ensure the physical conditions of the model are satisfied. The posterior allows for one or more consecutive reductions, fitting apparent features in the CMB power spectra in multipoles from a few tens to $\ell\simeq 2000$. As expected, these fits are not statistically favored with respect to the $\Lambda$CDM model. The methodology derived here allows for the inclusion of additional data sets (in particular, Large Scale Structure data), which in principle will increase the statistical significance of the reconstruction of the inflaton's speed of sound.
\end{abstract}

\maketitle

%\tableofcontents

%%%%%%%%%%%%%%%%%%%%%%%%%%%%%%%%%%%%%%%%%%%%%%%%%%%%%%%%%%%%%%%%%%%%%%%%%%
\section{Introduction}
%%%%%%%%%%%%%%%%%%%%%%%%%%%%%%%%%%%%%%%%%%%%%%%%%%%%%%%%%%%%%%%%%%%%%%%%%%
The standard cosmological model ($\Lambda$CDM) is currently favored by the available data. It assumes that primordial fluctuations are Gaussian and defined by an almost scale-invariant primordial power spectrum. These assumptions do not point to any particular origin, although the simplest inflationary model, canonical slow-roll single-field inflation, naturally predicts them. By contrast, other models of inflation predict deviations from the near scale-invariant spectrum in the form of features. If ever detected, they would open a new window of research in the field of primordial dynamics. See i.e: \cite{FeaturesReview, 2010AdAst2010E..72C, Slosar2019Scratches, Palma:2014hra}.

The study of features of primordial origin can be done within an Effective Field Theory approach. Within this scenario, features can be produced by the time dependence of primordial functions such as the slow-roll parameters or the speed of sound of the effective inflaton (the adiabatic mode). In particular, small, soft and transient reductions in the inflaton's speed of sound produce such correlated localized oscillatory features in the $n$-point correlation functions. In the 3-point function (or bispectrum), these localized oscillations present a distinct difference in phase between the squeezed and equilateral configurations \cite{2013PhRvD..87l1301A}.

The Planck Collaboration \cite{2018arXiv180706209P} searched for deviations of the canonical scenario in its last release of data. Nevertheless, they did not find strong evidence in the context of features in the primordial power spectrum \cite{2018arXiv180706211P}. They included, for the first time, a joint search of correlated simple features in the primordial power spectrum and in the bispectrum, also without significant results. However, the Planck Collaboration has not studied in detail different feature templates such as the above mentioned ones due to small and transient reductions of the inflaton's speed of sound. This motivates us to continue our previous study \cite{2014PhRvD..89j3006A, 2014PhRvD..90b3511A, Hu:2014hra, 2017PhRvD..96h3515T}, in preparation for a future release of the Planck bispectrum likelihood or for future investigation in light of incoming Large Scale Structure surveys.

Most of the time, the study of features in both observables (primordial power spectrum and higher correlation functions) are model-dependent, both regarding their physical origin and the ansatz used. In our latest paper in this series \cite{2017PhRvD..96h3515T}, we already pointed out the need for testing more flexible feature templates to mitigate the dependence on the ansatz. Within this approach, we can test whether multiple and consecutive reductions of the inflaton's speed of sound can take place consecutively, a possibility already pointed out in the previous work \cite{2017PhRvD..96h3515T}. Furthermore, reconstructing the inflaton's speed of sound allows us to test more complex feature templates with variable amplitude and oscillation frequency, which implies more possibilities to fit well-motivated deviations from $\Lambda$CDM beyond those that only used a pre-defined ansatz for the features.

Reconstructions at the level of the primordial power spectrum have already been attempted \cite{2016A&A...594A..20P, 2013JCAP...07..031H, 2014JCAP...01..025H, 2016JCAP...08..028R, 2018JCAP...02..012D, 2019PDU....23..245B, epsilon, 2019PhRvD.100j3511H}. However, there is not enough constraining power in Planck's power spectrum alone to decide on a particular model for the features. Model-informed reconstructions have the advantage of increasing the constraining power by adding the information contained in higher-order correlation functions; but this is only possible if the constraints of the theoretical model are properly imposed on the reconstructed spectrum, so that it will always lead to a consistent prediction. The task of imposing these physical constraints along the reconstruction is non trivial \cite{2016PhLB..760..297A}. It is advisable, instead, to reconstruct the primordial dynamics directly. In our case, we reconstruct the inflaton's speed of sound: the timing, intensity and rate of its reduction. Since we are reconstructing the underlying function leading to the correlated features, it is not only guaranteed that we will obtain a consistent bispectrum feature prediction using power spectrum data alone, but we will also be able to use both data sets simultaneously to get a more stringent reconstruction once a bispectrum likelihood has been released. 

In this paper, we develop a new analysis pipeline that uses Gaussian Processes (GPs), a hyper-parametric regression technique, to model the inflaton's speed of sound profile. The analytic nature of GPs makes easy to impose the constraints of the theoretical model, which involve derivatives of the reconstructed function. For a given number of nodes in the GP, we construct a prior on the hyper-parameters of the GP model (the position of the nodes and the correlation length), that maximizes entropy with respect to the bare physical constraints. In this way, we verify that nodes are not placed wherever they would lead to an unphysical reconstruction, and that the density with which the hyper-parameters are explored reproduces the measure of the physical prior \cite{2019Entrp..21..272H, Gariazzo:2018meg, Aghanim:2018eyx}. 

We test our new pipeline against Planck 2018 temperature, polarization and lensing CMB angular power spectrum data, obtaining corresponding posteriors of the parameters of interest and several \textit{maxima a posteriori}. Our results do not only reproduce our previous findings \cite{2017PhRvD..96h3515T}, but allow for combinations of multiple consecutive reductions as well as more complex shapes. 

This article is organized as follows. In section \ref{sec:theory} we review the theoretical framework for inflationary correlated features in the primordial power spectrum due to transient reductions in the speed of sound. In section \ref{sec:methodology}, we explain the methodology used to generate features in the primordial power spectrum: the parametrization for the reduction in the speed of sound (\ref{sec:methodology_c_s}), the chosen priors for the the different parameters (\ref{sec:methodology_priors}) and the computational procedure (\ref{sec:methodology_codes}). In section \ref{sec:results}, we present the results corresponding to the fitting of features using the CMB angular power spectrum. Finally, we discuss the results, draw our conclusion and show prospective work for the future in section \ref{sec:conclusion}.

%%%%%%%%%%%%%%%%%%%%%%%%%%%%%%%%%%%%%%%%%%%%%%%%%%%%%%%%%%%%%%%%%%%%%%%%%%
\section{Theoretical model}\label{sec:theory}
%%%%%%%%%%%%%%%%%%%%%%%%%%%%%%%%%%%%%%%%%%%%%%%%%%%%%%%%%%%%%%%%%%%%%%%%%%
We follow the Effective Field Theory (EFT) of inflationary perturbations \cite{2008JHEP...03..014C} to characterize the fluctuations of comoving curvature perturbations around an inflating cosmological background. It starts with an effective action for the Goldstone boson of cosmic time diffeomorphisms $\pi(t, \textbf{x})$. This Goldstone boson is related to the comoving curvature perturbation $\mathcal{R}(t, \textbf{x})$ through the relation $\mathcal{R} \approx -H(t)\pi(t, \textbf{x})$, with the Hubble parameter $H(t)\equiv \dot{a}/a$, with $a$ being the scale factor (where the dot denotes derivatives with respect to cosmic time $t$). The effective single field action for $\pi$ up to second order is given by
\begin{equation}\label{eq:S2}
S_2 = \int{d^4x a^3M_P^2\epsilon_1 H^2\left[-\frac{\dot{\pi}^2}{c_s^2}+\frac{(\partial_i\pi)^2}{a^2}\right]},
\end{equation}
where $M_P=1/\sqrt{8\pi G}$ is the reduced Planck Mass in natural units $c= \hbar = 1$, $\epsilon_1 \equiv  -{\dot{H}}/{H^2}$ is the first slow-roll parameter and $c_s\equiv c_s(t)$ is the time-dependent speed of sound. \\

The effective single field action up to third order, neglecting higher order slow-roll corrections ($\sim \mathcal{O}(\epsilon_1^2)$) and assuming $\dot{\pi}^3$ to be small and approximately constant reads \cite{Achucarro:2012sm},
\begin{align}\label{eq:S3}
    S_3 = & \int{ \bigg [ d^4x a^3M_P^2\epsilon_1 H^2} \nonumber \\
     &{}  -2Hsc_s^{-2}\pi\dot{\pi}^2- (1-c_s^{-2})\dot{\pi}   \left(\frac{\dot{\pi}^2}{c_s^2}- \frac{(\partial_i\pi)^2}{a^2}\right)  \bigg] ,
\end{align}
where $s\equiv s(t)$ parameterizes the change in the speed of sound $c_s(t)$ defined as
\begin{equation}\label{eq:s}
s \equiv \frac{\dot{c_s}(t)}{(c_s(t) H)}.
\end{equation}
The physical details of the theory are encoded in the speed of sound $c_s$ and in its corresponding rate of change given by $s$. The speed of sound $c_s$ accounts for the effects of integrating out the heavy fields within the effective action. To get an insight of what this variable $c_s(t)$ means, we look at the particular case of an effective theory for the comoving curvature perturbation $\mathcal{R}$, when a strong turn in the inflationary trajectory in multifield space is supported by a heavy field $\mathcal{F}$ with “effective mass” $M_{\text{eff}}$. In this case, the curvature perturbation $\mathcal{R}$ is kinetically coupled to the heavy field $\mathcal{F}$. This effective action is similar to the EFT of inflation equation~(\ref{eq:S2}), with the speed of sound $c_s$ of the adiabatic perturbation $\mathcal{R}$ given by \cite{2011JCAP...01..030A, 2011PhRvD..84d3502A}
\begin{equation}\label{eq:c_s}
c_s^{-2}=1+\frac{4\Omega^2}{k^2/a^2+M_{\text{eff}}},
\end{equation}
where $\Omega$ is the the angular velocity when there is a turn in the inflationary trajectory, inducing a momentary reduction on the speed of sound $c_s$ \cite{Achucarro:2012sm}. The effect of this variable speed of sound $c_s$ can be seen in the primordial power spectrum $\mathcal{P_R}$, in the bispectrum $\mathcal{B_R}$ and in higher-order correlation functions. In particular, transient variations of $c_s$ produce localized oscillatory and correlated features in both  $\mathcal{P_R}$ and  $\mathcal{B_R}$ \cite{EFTCS1}. Generally, $c_s(t)$ encodes the effect of derivative interactions.

%----------------------------------------------------%
%\subsection{Features in the Primordial Power Spectrum $\mathcal{P_R}$}\label{sec:theoretical_model_PPS}
The almost scale-invariant featureless power spectrum $\mathcal{P_{R}}_0$, with $c_s = 1 \rightarrow u = 0$, is defined as,
\begin{equation}\label{eq:PPS}
    \mathcal{P_{R}}_0 = A_s\left(\frac{k}{k_*}\right)^{n_s-1},
\end{equation}
where $A_s$ is the scalar amplitude, $k_*$ is the pivot scale and $n_s$ the so-called spectral index, which depends on the slow-roll parameters as,
\begin{equation}
    n_s \equiv 1-2\epsilon_1-\epsilon_2,
\end{equation}{}
where $\epsilon_2 \equiv \dot{\epsilon_1}/(\epsilon_1 H)$ is the second slow-roll parameter.
Under the assumption of small, mild and transient reductions of the speed of sound $c_s$, the modifications in the primordial power spectrum of curvature perturbations $\Delta \mathcal{P_R} /\mathcal{P_{R}}_0$ were already calculated \cite{2013PhRvD..87l1301A}. The quadratic action of EFT of inflation, equation~(\ref{eq:S2}), is divided into a free part (resembling single field inflation, with $c_s=1$) and a small perturbation:
\begin{align}\label{eq:S2_perturbed}
    S_2 = &\int \mathrm{d}^4 x \, a^3 M_{\mathrm{P}}^2 \epsilon H^2 \left( \dot\pi^2 - \frac{\left( \partial_i \pi \right)^2}{a^2} \right) \nonumber \\
    &{} - \int \mathrm{d}^4 x \, a^3 M_{\mathrm{P}}^2 \epsilon H^2 \bigg( \left( 1 - c_s^{-2} \right) \dot\pi^2 \bigg) ,
\end{align}
Transitioning from cosmic time $t$ to the conformal time $\tau$, so that $d\tau=dt/a(t)$, using the \textit{in-in} formalism \cite{2005PhRvD..72d3514W} and the following definition of the variable $u$,
\begin{equation}\label{eq:u_first}
u(\tau)\equiv(1-c_s^{-2}(\tau)), 
\end{equation}
the change in the primordial power spectrum $\Delta \mathcal{P_R}$ is given by the Fourier transform of the reduction in the speed of sound $c_s$:
\begin{equation}\label{eq:deltaPPS}
\frac{\Delta \mathcal{P_R}}{\mathcal{P_{R}}_0} = k\int_{-\infty}^0{d\tau u(\tau)\sin{(2k\tau)}}.
\end{equation}

%-------------------------------------%-------------------------------------%-------------------------------------
\section{Methodology}\label{sec:methodology}

\begin{figure}[htp]
  \centering
  \includegraphics[width=1.0\columnwidth]{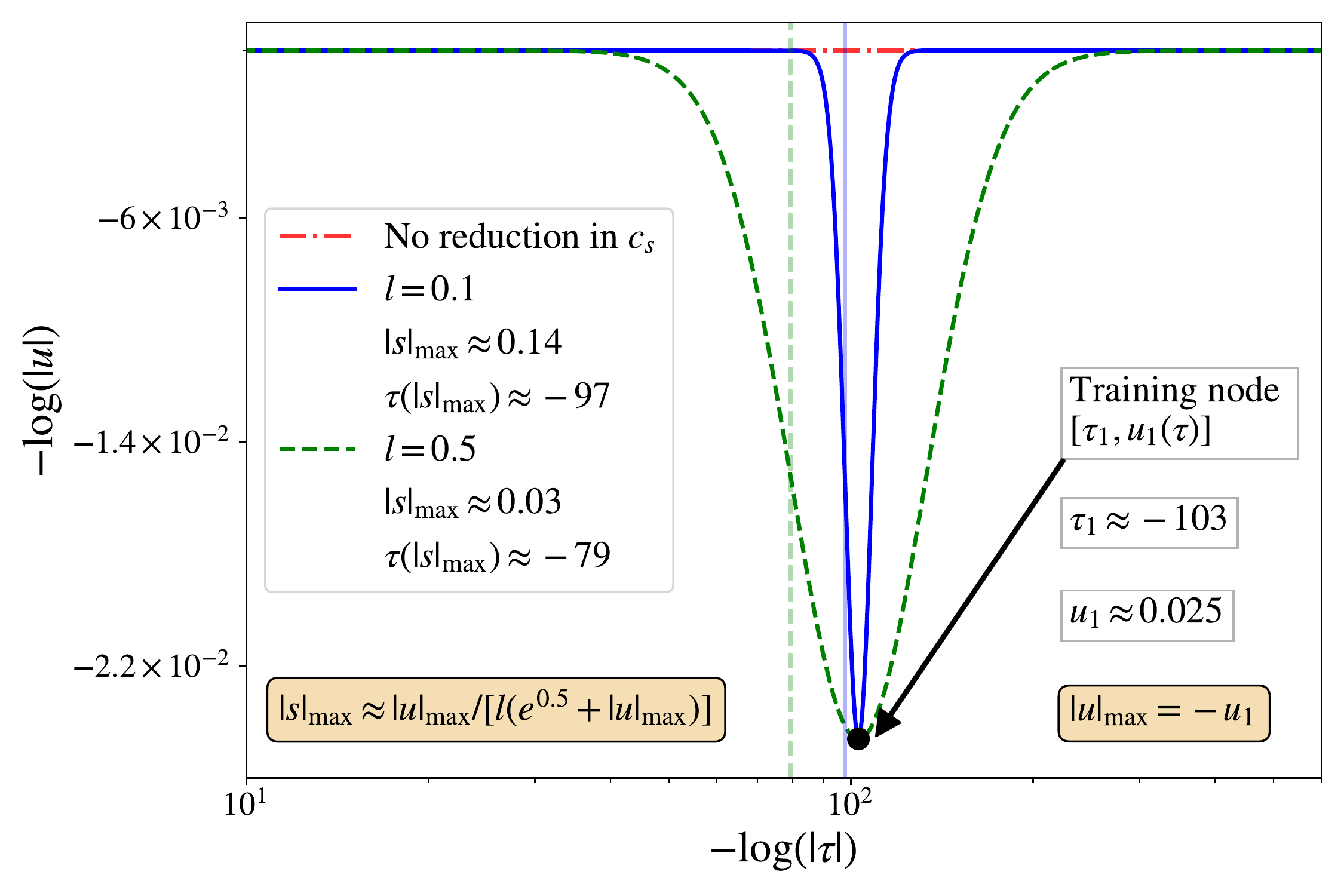}
  \caption{Example of reductions of the inflaton's speed of sound $u(\tau)$ in logarithmic space GP with a single node at $(\tau_1, u_1)$. The reductions peak at $\tau_1 \approx 103$ with a maximum reduction value $-u_1=|u|_\mathrm{max}\approx 0.025$. The width of each of the reductions (given by the correlation length $l$, which, for a single GP node, plays the role of the standard deviation) is different, being the green-dashed parametrization ($l=0.5$) milder than the solid blue one ($l=0.1$). The rate of change of $u(\tau)$, see Eq.\ \eqref{eq:s}, can be approximated as $|s|_\mathrm{max} = |u|_\mathrm{max}/[l(e^{0.5}+|u|_\mathrm{max})]$ using this parametrization. The vertical lines indicate the values of $\tau$ at which $|s|_\mathrm{max}$ is reached in each case.}
    \label{fig:explanatory}
\end{figure}

The reduction of the speed of sound and its rate of change are encoded in $u(\tau)$ and $s(\tau)$ respectively. We aim to use current cosmological data (the temperature, polarization and lensing power spectrum of Planck 2018) to estimate them given the theoretical framework presented in section \ref{sec:theory}. To do that, we use \textit{Bayesian inference}. The estimation of the joint probability distribution of a set of parameters $\theta$ of a model $M$ given some data $d$, the so-called posterior $P(\theta|d, M)$, is computed using Bayes' Theorem \cite{bayesian}:
\begin{equation}\label{eq:parameter_estimation}
P(\theta|d, M) \propto \mathcal{L}(d|\theta, M) \Pi(\theta|M),
\end{equation}
where $\mathcal{L}(d|\theta, M)$ is the \textit{likelihood} (probability of observing the data $d$ given the model $M$ is realized with parameters $\theta$) and $\Pi(\theta|M)$ the \textit{prior} (probability distribution of the parameters $\theta$ given some \textit{a priori} information). In this section, we present the methodology that we have employed to study the posterior distribution of $u(\tau)$, and hence $s(\tau)$.

\subsection{Reconstruction model for the reduction in the speed of sound}\label{sec:methodology_c_s}

\begin{figure*}[ht]
  \centering
  \includegraphics[width=1.00\textwidth]{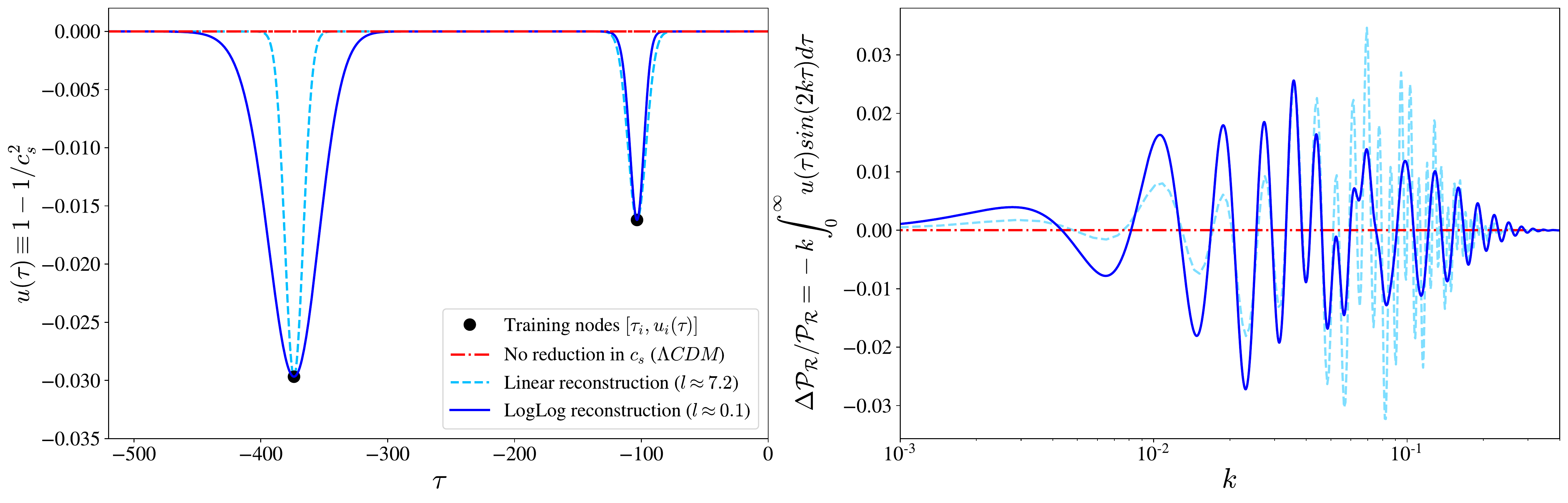}
  \caption{
      \textbf{Left panel}: Reconstruction of $u(\tau)$ using a GP on $\log|u(\log|\tau|)|$ (continuous line) and on $u(\tau)$ (dashed line), both using as training nodes $(\tau, u) \in \left\{(-100, -0.016), (-375, -0.03)\right\}$, and correlation length for each model such that the width of the mode corresponding to the first node is similar (notice the difference in width of the mode of the second one). As the reconstruction is done in conformal time $\tau$, the x-axis is always negative and $\tau = 0$ indicates the end of inflation.
      \textbf{Right panel}: Corresponding feature in the primordial power spectrum. Notice how, despite the similar position of the training nodes, their features look quite different: similar in the leftmost oscillations (corresponding to the node at $\tau=-100$), but very different after that, due to the broader width of the first mode.}
    \label{fig:pipeline}
\end{figure*}

We use Gaussian Processes (GPs) \cite{ML} as an interpolator for reconstructing the speed of sound of the inflaton. The \textit{mean} curve of a GP, which we use to represent the speed of sound evolution, is smooth by construction (given our choice of kernel, see below) and naturally returns to a baseline value away from the \textit{nodes} of the interpolator, which is useful for representing the transient character of the speed of sound reductions. The length scale over which the return to the baseline happens is called \textit{correlation length}, and it is the same for all individual nodes in a particular realisation. The particular properties of the correlation between nodes is given by the \textit{kernel} of the GP, which we choose to be a \textit{squared exponential} kernel. This means that when the interpolator is defined by a single node placed at $(x_1, y_1)$, the interpolating curve looks like a (non-normalised) Gaussian peaking at the node's position, with standard deviation equal to the correlation length $l$, i.e.\ $y(x)=y_1\ \exp\left[-1/2\ (x-x_1)^2/l^2\right]$.

We aim to reconstruct $u(\tau)\equiv(1-c_s^{-2}(\tau))$. Since $u(\tau)$ is a negative quantity, it makes sense to reconstruct the logarithm of $-u=|u|$, to guarantee that the GP interpolator, once exponentiated, conserves sign. On the other hand, there is a choice to be made about the scale of the conformal time axis: whether to reconstruct $\log|u(\tau)|$ or $\log|u(\log|\tau|)|$. We show results for the latter case in this section for illustration purposes.

As explained above, in the case where a single node is placed at $(\tau_1, \log(-u_1))$ the reconstruction of $\log(-u(\tau))$ corresponds to a single transient reduction given by a log-normal function of conformal time, whose maximum occurs exactly at the node. The parameters that we would try to infer from the data would be the position of the node $(\tau_1, \log(-u_1))$ and the correlation length $l$ representing the standard deviation of the log-normal. The rate of change of the reduction, of interest in our theoretical framework, would peak approximately at $|s|_\mathrm{max}\approx0.5l |u_1|/(\exp(0.5)+|u_1|)$. See Fig.\ \ref{fig:explanatory} for an example.

To reconstruct $\log|u(\log|\tau|)|$ as a generalization of the previous case using GPs, we choose a number $i$ of training nodes\footnote{Notice that our use of GPs as interpolators does not involve machine-learning, but we are borrowing the term \textit{training} from its literature.} $(\log|\tau_{i}|, \log|u_{i}|)$ where $\log|u_{i}|\defeq\log|u(\log|\tau_{i}|)|$ (see Fig.\ \ref{fig:pipeline}), to which we fit a GP with kernel function 
\begin{multline}\label{eq:u_tau}
    \kappa(\log|\tau_i|, \log|\tau_{i+1}|;\,l) = \\ c^2 \exp\left\{-\frac{1}{2}\left(\frac{\log|\tau_i| - \log|\tau_{i+1}|}{l}\right)^2\right\} ,
\end{multline}
where $c$ is the \emph{output scale}, and $l$ the \emph{correlation length}. The output scale $c$ plays no role in this approach, and can be fitted using \emph{maximum likelihood} and then ignored. The correlation length will be sampled together with the position of the nodes. To compute the GPs, we use the Python package \texttt{sklearn} \cite{sklearn}. The \emph{mean} of the GP is used as an interpolator for $\log|u(\log|\tau|)|$, and reads, in terms of the training nodes, as the matrix product:
\begin{equation} \label{eq:gppred}
\begin{split}
  \log|u(\log|\tau|)| = & \kappa(\log|\tau|, \log|\tau_i|;\,l) \times \\
                        & \left[\kappa(\log|\tau_i|, \log|\tau_j|;\,l)\right]^{-1} \log|u_j|,
  \end{split}
\end{equation}
where the first kernel function $\kappa$ is a vector of evaluations at the requested $\log|\tau|$ combined with each of the training $\log|\tau_i|$, the second one is the matrix of evaluations of $\kappa$ for each pair of training nodes $(i,j)$, and the final term is the vector of training $\log|u_j|$. Once $u(\tau)$ is generated, we calculate $s(\tau)$ from Eq.\ \eqref{eq:s} numerically, which we can rewrite more conveniently as 
\begin{align}\label{eq:s_tau}
  s(\tau) & =  \frac{1}{2} \frac{u}{1-u} \frac{\diff \log u}{\diff \log|\tau|} \nonumber\\
          & = \frac{1}{2} \frac{u}{u-1} \frac{1}{l^2}
             \left[\left(\log|\tau| - \log|\tau_i|\right) \kappa(\log|\tau|, \log|\tau_i|;\,l)\right] \times \nonumber \\
          & \phantom{=} \left[\kappa(\log|\tau_i|, \log|\tau_j|;\,l)\right]^{-1} \log|u_j|,
\end{align}
where we have taken the derivative after substituting $u$ by the mean of the GP defined in Eq.\ \eqref{eq:gppred}. Notice that this reproduces the matrix product in Eq.\ \eqref{eq:gppred}, just changing the first vector. Finally, we compute the power spectrum feature of Eq.\ \eqref{eq:deltaPPS} from a fine sampling of the GP using the \texttt{FFTLog} algorithm \cite{PaperFFTLOG, pythonwrapperFFTLOG}. The density and limits of the $\log|\tau|$ sampling for the \texttt{FFTLog} are chosen adaptively to minimise computational costs and guarantee the accurate computation of the transform.

The most consequential difference of the choice between linear and logarithmic $\tau$ in the GP will show up whenever we have nodes separated by a distance much larger than the correlation length, appearing as isolated (log)Gaussians: in the linear case, their width in $\tau$ will be similar, whereas for the logarithmic one, the width will scale logarithmically (see Fig.\ \ref{fig:pipeline}). Looking at the first equality in Eq.\ \eqref{eq:s_tau}, and seeing how $s$ depends on the logarithmic derivative on $\tau$, it is easy to see that the linear parameterization is going to struggle to place two or more nodes away from each other, since $s(\tau)$ will peak at highly different values in each of them, making it hard not to violate the perturbativity bounds on $s$ (see sec.\ \ref{sec:methodology_priors}).

Thus for the primary results in this paper, we model $\log|u(\log|\tau|)|$. To mitigate excessive sensitivity to the prior of our results, we also perform a reconstruction in $u(\tau)$. Notice that by modelling $u$ and not $\log|u|$ we need to deal with cases in which $u(\tau)$ goes positive, by assigning it null prior density. However, those are generally disfavored by the data (require large $l$ compared to the distance between nodes), and the large difference between the $\log|u(\log|\tau|)|$ and the $u(\tau)$ reconstruction is useful for assessing prior sensitivity.

\subsection{Parameters and Priors}\label{sec:methodology_priors}

\begin{figure*}[ht]
    \centering
    \includegraphics[width=1.0\textwidth]{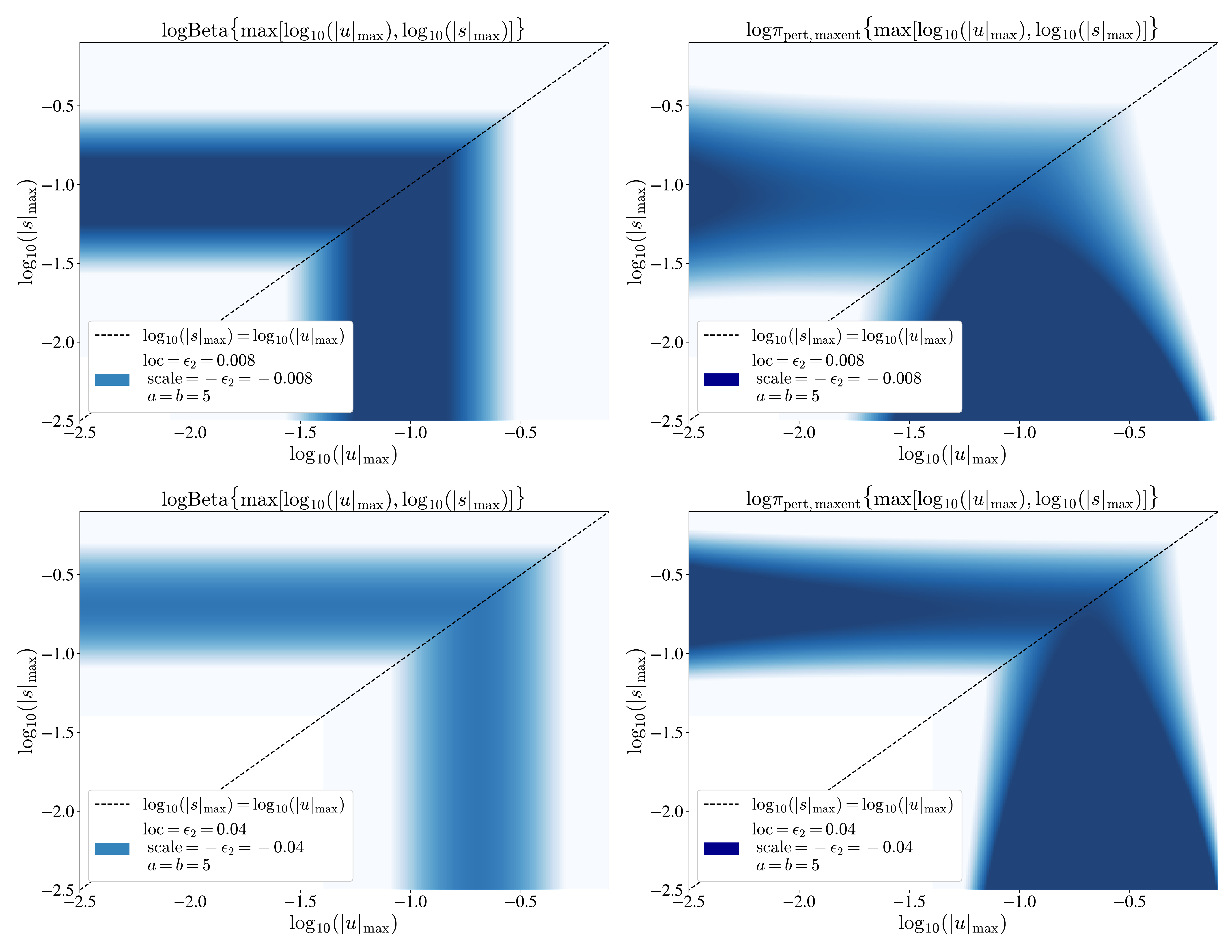}
    \caption{A description of the Bayesian priors adopted in this study. \textbf{Left panels}: Desired Beta density distribution for the parameters $log_{10}(|u|_\mathrm{max})$ and $log_{10}(|s|_\mathrm{max})$, with $\epsilon_2=0.008$ and $\epsilon_2=0.04$, respectively. \textbf{Right panels}: probability density distribution induced by $\pi_0$ on $log_{10}(|s|_\mathrm{max})$ and $log_{10}(|u|_\mathrm{max})$ (defined as $P$ in the denominator of Eq.\ \eqref{eq:maxent}), for a Gaussian-Processes reconstruction in $\log|u(\log|\tau|)|$ using 2 training nodes. The initial prior $\pi_0 (\tau_i, u_i, l)$ induces overdensities at low $log_{10}(|s|_\mathrm{max})$ due to this quantity being doubly-correlated to early-time position of the nodes and large correlation lengths; Eq.\ \eqref{eq:maxent} corrects for this effect.}
    \label{fig:perturbativity}
\end{figure*}

The action described in equations~(\ref{eq:S2}) and~(\ref{eq:S3}) is perturbative in terms of $(1-1/c_s^2)$. It implies that the reduction in the speed of sound, $c_s$, cannot be too big ($|u|\ll 1$) and the rate of change in the reduction cannot be too fast ($|s|\ll 1$). Also, the contributions of the slow-roll corrections $\epsilon_1, \epsilon_2$ have to be smaller than those of the variable speed of sound $c_s$. We need to impose these conditions for all values of $\tau$, but it is enough to restrict to the point where $u(\tau)$ and $s(\tau)$ take their maximum value $(|u|_{\text{max}}, |s|_{\text{max}})$. Note that imposing the perturbative limit on $|s|_{\text{max}}$ satisfies the consistency conditions in \cite{EFTCS2, 2014PhRvD..89h3531A, 2014PhRvD..89l7301C}. In short:

\begin{equation}\label{eq:condition}
\text{max}(\epsilon_1, \epsilon_2) \ll \text{max}(|u|_{\text{max}}, |s|_{\text{max}})\ll 1.
\end{equation}

In \cite{2017PhRvD..96h3515T} we argued that this condition could be naturally imposed by a prior $\mathrm{Beta(5,5)}$ on $\text{max}(\log_{10}|u|_{\text{max}}, \log_{10}|s|_{\text{max}})$ between the extremes in Eq.\ \eqref{eq:condition} (see left part of Fig.\ \ref{fig:perturbativity}), the logarithm coming from the difference in order of magnitude between both bounds.
  
Contrary to \cite{2017PhRvD..96h3515T}, in this work $|u|_{\text{max}}$ and $|s|_{\text{max}}$ are not sampled directly. Instead, the parameter space for the feature consists of the position of the nodes $\left\{\left(\tau_i, u_i=u(\tau_i)\right)\right\}$ and the correlation length $l$. Thus, the total number of feature parameters is $2N+1$ for a number $N$ of nodes. Imposing the $\mathrm{Beta}$ prior described above is not as simple as sampling the GP parameters from some prior, computing $|u|_{\textnormal{max}}$ and $|s|_{\textnormal{max}}$ along the reconstruction, and multiplying by the $\mathrm{Beta}$ density. That procedure will likely introduce undesired information in the shape of under- or overdensities in the probability induced on $(|u|_{\textnormal{max}}, |s|_{\textnormal{max}})$, which would finally diverge significantly from a $\mathrm{Beta}$. The correct way to proceed so that the induced probability on the physical parameters is the desired one is by constructing a distribution on the parameters of the nodes that \emph{maximises entropy} with respect to the desired one, which can be computed, according to \cite{2019Entrp..21..272H}, as
\begin{equation}\label{eq:maxent}
  \pi_\text{pert, maxent}(\tau_i, u_i, l) = \frac{\pi_0(\tau_i, u_i, l) \pi_\text{pert}(|u|_\text{max}, |s|_\text{max})} {P(|u|_\text{max}, |s|_\text{max}\textnormal{   }|\textnormal{   }\pi_0)} ,
\end{equation}
where $\pi_0(\tau_i, u_i, l)$ is some initial prior on the node parameters and $\pi_\text{pert}(|u|_\text{max}, |s|_\text{max})$ the $\mathrm{Beta}$ prior described above. The term in the denominator is the probability density induced by $\pi_o$ on $(|u|_\text{max}, |s|_\text{max})$, which we compute from a Monte Carlo sample from $\pi_0$ using \texttt{PolyChord} \cite{Polychord, Polychord2}. The Monte Carlo sample is fed to \texttt{GetDist} \cite{getdist} to construct a density estimator. Both the $\mathrm{Beta}$ prior and the maxent prior can be seen in Fig.\ \ref{fig:perturbativity}.

For $\pi_0$, we choose log-uniform prior distributions on the correlation length $l$ and on the training node location $(\tau_i, u_i)$\footnote{Note that we are sampling the training nodes and the correlation length in a logarithmic scale as we expect them to vary several orders of magnitude.}. The bounds for the time-positions $\tau_i$ are chosen so that the feature falls in the CMB window function (features running from scales $k\approx10^{-3}$ to  $k\approx3\times10^{-1}$, though a larger region has been scanned as a consistency check (see Sec.\ \ref{sec:checks}). The bounds for the amplitude of the reductions at the nodes, $u_i$, are chosen to generously fulfill the EFT condition in Eq. \eqref{eq:condition}. Summarizing:
\begin{align}\label{eq:priors}
     \pi_0(\log_{10}(|\tau_1|), \ldots, \log_{10}(|\tau_n|), \log_{10}(|u_1|), \ldots,  & \nonumber \\ \log_{10}(|u_n|), \log_{10}l) =  \mathcal{U}(1.8 < \log_{10}(|\tau_n|) < 3.3) \times \nonumber \\ \prod_{i=n}^{2} \mathcal{U}(\log_{10}(|\tau_i|) < \log_{10}(|\tau_{i-1}|) < 3.3) \times \nonumber & \\
     \prod_{i=1}^{n} \mathcal{U}(-4 < \log_{10}(|u_i|) < 0)\times \pi(l)
\end{align}
where $\mathcal{U}$ means a uniform distribution, and the prior on the time positions of the nodes includes sorting so that $\tau_i < \tau_{i-1}$ ($i$ runs from $1 \ldots n$). The prior on $l$ is chosen so that it produces reasonable values of $|s|_{\text{max}}$. For each of the two reconstructions studied here, $\log|u(\log|\tau|)|$ and $u(\tau)$, the boundaries can be chosen as\footnote{Notice that, while smaller values of $l$ will result in sharper reductions with too high, forbidden $|s|_{\text{max}}$ values, larger values of $l$ would result in small $|s|_{\text{max}}$ values which are actually allowed as long as $|u|_{\text{max}}$ fulfills Eq.\ \eqref{eq:condition}. In any case, we are imposing these upper $l$ boundaries for the main runs, since reductions with very small $|s|_{\text{max}}$ tend not to be easily distinguishable from changes in the background cosmological model (see Sec.\ \ref{sec:checks}).}
\begin{align}
  & \pi_{\log|u(\log|\tau|)|}(l) = \mathcal{U}(-2 < \log_{10}l < 2) \qquad\text{and}\qquad \nonumber \\
  & \pi_{u(\tau)}(l) = \mathcal{U}(-2 < \log_{10}l < 3.3).
\end{align}
As an improvement on \cite{2017PhRvD..96h3515T}, in this work we do not fix the value of the slow-roll parameters $\epsilon_1$ and $\epsilon_2$ as bounds of the perturbativity condition Eq.\ \eqref{eq:condition}. Instead we let the bounds of the $\mathrm{Beta}$ distribution run dynamically, marginalizing over the slow-roll parameters. We use as priors uniform distributions $\mathcal{U}(0.0001< \epsilon_1 < 0.05)$ and $\mathcal{U}(-0.06< \epsilon_2 < 0.06)$, which encompass the $\Lambda$CDM posterior found for them in Planck 2018 \cite{2016A&A...594A..20P}.

\begin{figure}[ht]
    \centering
    \includegraphics[width=1.0\columnwidth]{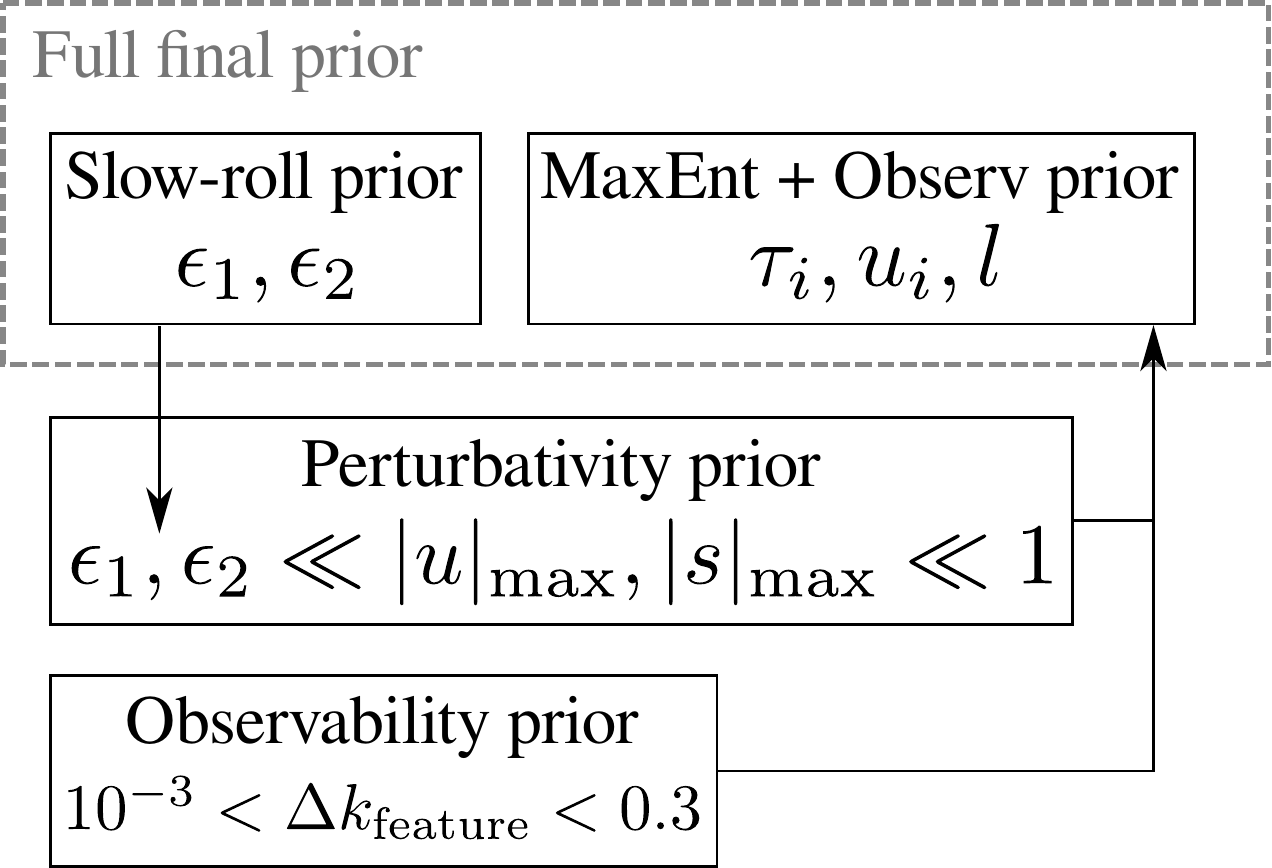}
    \caption{Diagram showing the structure of the prior. The conformal time $\tau_i$ of the nodes must fulfil that the feature happens within the observable CMB window. The physically-motivated perturbativity condition of Eq.\ \eqref{eq:condition} is imposed using \textit{maximum entropy} (see text) on the position of the nodes and the correlation length. Since the value of the slow-roll parameters influences the prior on the position of the nodes via the lower bound of the perturbativity condition, the full prior (dashed grey box) is non-separable.}
    \label{fig:priordiag}
\end{figure}

A diagram showing the different assumptions that enter the final prior is shown in Fig.\ \ref{fig:priordiag}.

%-------------------------------------
\subsection{Data sets and sampler}\label{sec:methodology_codes}

To constrain the reduction of the speed of sound, we use the Planck 2018 polarized CMB and lensing data. In particular we use the product of the low multipole likelihoods \texttt{lowT} and \texttt{lowE}, the \emph{unbinned} high-$\ell$ likelihood \texttt{plik\_TTTEEE} and the \texttt{lensing} likelihood. We use the unbinned likelihoods because of the fast frequency of oscillations in the features, as was already pointed out in \cite{2017PhRvD..96h3515T}.

We compute the changes to the CMB power spectra $C_\ell$ using the Boltzmann code \texttt{CAMB} \cite{Lewis:2013hha}, modified accordingly to account for the increased sampling in $k$ needed by the oscillatory features $\Delta \mathcal{P_R}/\mathcal{P_{R}}_0$ in the primordial power spectrum. We sample over the parameter space described in section \ref{sec:methodology_priors}, i.e.\ the positions of the training nodes $\{(\tau_i, u_i=u(\tau_i))\}$, the correlation length $l$ of the GP, and the kinetic slow-roll parameters $\epsilon_1$ and $\epsilon_2$. We also allow for the possibility of tensor modes, as changes in the Sachs-Wolfe plateau caused by them could possibly be correlated with features at very large scales. We track as derived parameters the scalar tilt $n_s$, the tensor-to-scalar ratio $r$ and the EFT parameters $(|u|_\text{max}, |s|_\text{max})$. We fix the rest of cosmological parameters of $\Lambda$CDM to the best fit of Planck 2018 with the present likelihoods, as well as the \emph{nuisance} parameters of the likelihoods. Fixing the $\Lambda$CDM parameters is justified by previous sensitivity analyses in \cite{2017PhRvD..96h3515T}, that we repeat here for the background $\Lambda$CDM parameters by exploring a broader range of $\tau$ and $l$ than the one indicated above (see Sec.\ \ref{sec:checks}, where we have also assessed the impact of fixing the nuisance parameters of the Planck likelihoods).

We obtain the posterior distribution of the parameters using the sampler \texttt{PolyChord} \cite{Polychord, Polychord2}. We use this nested sampler since, from previous searches, we expect the posterior distributions of $u_{\text{max}}$ and $\tau_i$ to be multi-modal. The handling of the priors, likelihoods, Boltzmann code and sampler is managed by the Bayesian framework \texttt{Cobaya} \cite{2020arXiv200505290T}. The analysis of the posterior distributions is carried out using \texttt{GetDist} \cite{getdist}.

We sample the posterior of two different parameterizations of the GP sound speed reconstruction: $\log|u|(\log|\tau|)$ and $u(\tau)$, in the following called simply \emph{logarithmic} and \emph{linear} parameterizations, respectively. We know the logarithmic parameterization is more stable numerically, as it consistently makes the reconstruction of $u(\tau)$ negative. However, we still use the runs in the linear parameterization for the purposes of assessing prior sensitivity. For the first sampling processes (up to three GPs nodes), we run \texttt{Cobaya} in parallel launching 8 MPI processes, each allowed to thread across 3 CPU cores. In the case of 4 nodes, we run \texttt{Cobaya} with 32 MPI processes, each allowed to thread across only one single CPU core. The nested sampler \texttt{PolyChord} has been run with 1000 live points (which is far above the requirements for the current number of dimensions in the parameter space) and a stopping criterion of 0.01. The computation time varies depending on the number of training nodes in the GPs: from a few days with only 1 node, up to several weeks with 4 nodes. 

All the \textit{maxima a posteriori} (MAP) presented in the next section have been obtained running \texttt{Py-BOBYQA} \cite{2018arXiv180400154C, 2018arXiv181211343C} (a Python implementation of the \texttt{BOBYQA} algorithm \cite{BOBYQA}, available via \texttt{Cobaya}), initialized on the relevant local maxima of the \texttt{PolyChord} samples.

\section{Results}\label{sec:results}

\subsection{Consistency checks}\label{sec:checks}

Before presenting our results, we shortly discuss whether the assumptions made in previous sections were justified. In particular, we have tested whether we find clear posterior modes outside the $(\tau_i, l)$ prior region described section \ref{sec:methodology_priors} (the \textit{CMB window} prior), and whether in posterior modes either in our initial prior or in the broader region, the assumption of no-correlation with background cosmological parameters is fulfilled. 

To do that, we produced a 1-node posterior sample in the logarithmic parameterization in the enlarged prior region $0<\log_{10}(|\tau_i|)<4.3$ and $2<\log_{10}l<10$, and let the background $\Lambda$CDM parameters vary. No significant modes were found outside the original, reduced prior region. We found mild modes in the region $1.8< \log_{10}(|\tau_i|)<3.3$ and $2 <\log_{10}l<3.3$, which presented some degeneracy between $\Omega_m$, 
$n_s$ and the reconstruction parameters  ($\rho \approx 0.17$), due to the fact that these features can be confused with the shape of the first and second acoustic peaks (already observed in \cite{2017PhRvD..96h3515T}). This justifies restricting ourselves to the prior described in \ref{sec:methodology_priors}, since any mode found outside of it would not be distinguishable from background cosmology.

\begin{figure}[ht]
    \centering
    \includegraphics[width=0.8\columnwidth]{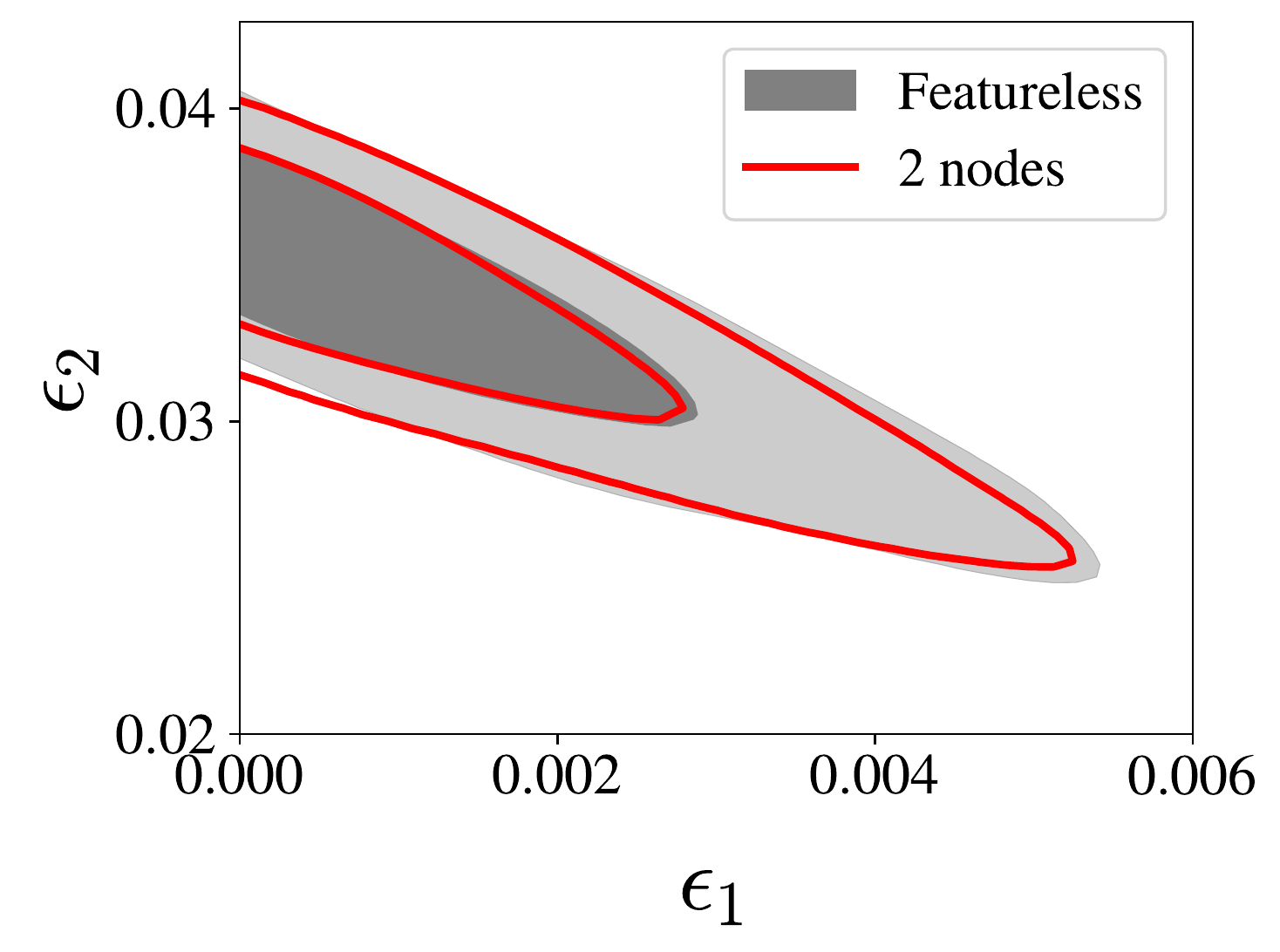}
    \includegraphics[width=0.8\columnwidth]{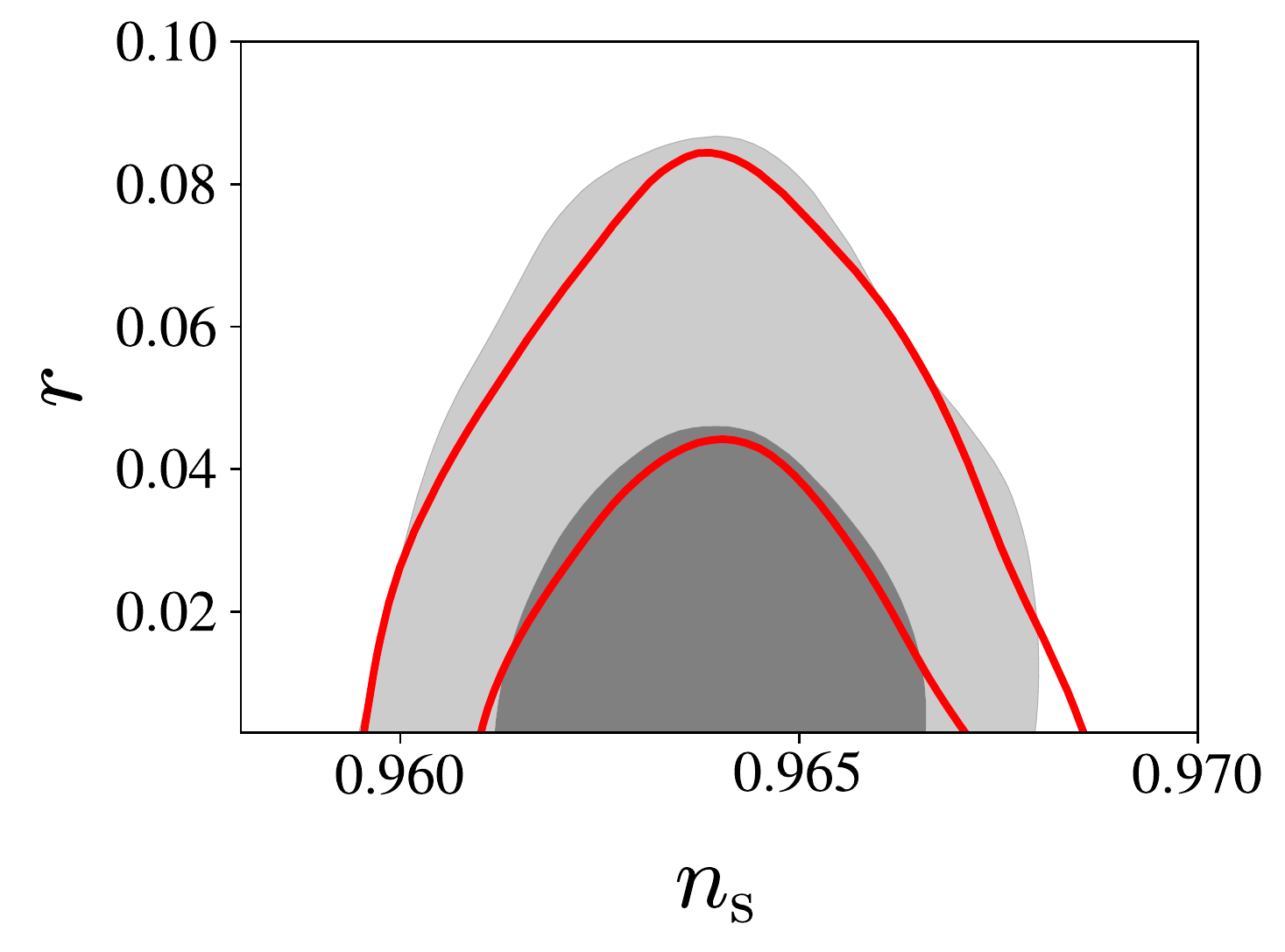}
    \caption{Posterior distributions of the primordial parameters: the kinetic slow-roll parameters $\epsilon_1$ and $\epsilon_2$, and the derived spectral index $n_s$ and tensor-to-scalar ratio $r$, for a 2-node reconstruction in the logarithmic parameterization (red line). The grey contours correspond to the featureless $\Lambda$CDM scenario. The correspondence between both posteriors is due to the absence of degeneracies between the $c_s$ reconstruction parameters and the slow-roll parameters. Similar results are found for 1, 3 and 4 nodes.}
    \label{fig:2nodes_eps12}
\end{figure}

The check for degeneracies between the $c_s$ reconstruction parameters and the slow-roll parameters is of particular importance, since the latter determine the perturbative prior limits on the former (see Eq. \ref{eq:condition} and Fig.\ \ref{fig:perturbativity}). We have found no significant degeneracies, neither in the tests described above nor in the final runs. We have reproduced the Planck $\Lambda$CDM posterior on the slow-roll parameters in all cases (see Fig.\ \ref{fig:2nodes_eps12}).

Most of the results below have been run both in the linear and logarithmic parameterizations for the Gaussian Process (GP) reconstruction of the speed of sound profile $u(\tau)$. The results agree with each other, in particular, for the maxima a posteriori found at late conformal time (i.e.\ towards the end of inflation, with $-\tau_i$ of a few hundreds, where both parameterizations look similar). However, the logarithmic parameterization differs from the linear one when training nodes are thrown at early values of conformal time (i.e. $-\tau_i$ over 800), see Fig.\ \ref{fig:pipeline}). This is due to modes of constant width in logarithmic scale getting broader the further we go along the axis of conformal time. In the 1-node case, the linear parameterization reproduces the results in \cite{2017PhRvD..96h3515T} (which uses a Gaussian ansatz in $u(\tau)$), whereas the logarithmic parameterization produces different 1-node posterior modes (see Sec. \ref{sec:1mode} in appendix A).

It is worth remarking that the use of the logarithmic parametrization does not compromise the flexibility of our reconstruction of $u(\tau)$. Even though the logarithmic parameterization reconstructs naturally profiles of $u(\tau)$ with broad reductions at earlier conformal times and narrower reductions at later times (which is preferred so that $|s|_{\textnormal{max}}$ is not violated), narrow reductions at early conformal times can always be achieved by adding further nodes that would force the profile to return to zero. If the data and EFT conditions did allow for a narrow reduction at earlier conformal times, we would have seen it during the analysis of the posterior distributions when more than one training node was used.

Moreover, we have assessed the effect of fixing the nuisance parameters of the Planck likelihoods by running a minimizer around both the baseline, featureless $\Lambda$CDM model and the MAP features presented below, now letting the nuisance parameters vary. We find that this choice has almost no impact in our analysis, changing the $\Delta\chi^2$ values by less an unit.

Finally, we have assessed the impact of using separately each of the high-$\ell$ \textit{unbinned} TT and EE Planck 2018 data sets, in order to check which subset dominates the posterior around each of the fits. To do that, we have combined each of these subsets with low-$\ell$ temperature and polarisation, and lensing data, fixing LCDM and nuisance parameters, and assuming a single dip. Using high-$\ell$ TT data alone, we recover the main single-dip maxima a posteriori described in the next section and appendix \ref{sec:appendix_A} (i.e:\ $\tau_1 \approx -100, -200, -400, -1000$). When using high-$\ell$ EE data alone, we clearly recover the dip at $\tau_1 \approx -100$, whereas just barely the peaks at $\tau_1 \approx -200$ and  $\tau_1 \approx -400$, and none of the earlier-time dips found in combination with TT data. This shows that our posterior is temperature-dominated in most of the parameter range. Concordance at late times is encouraging, but it needs to be explored further with future less noisy polarised data from space- and ground-based surveys.

\subsection{Reconstruction of the inflaton's speed of sound profile $u(\tau)$}

In this section, we present the results of the GP reconstruction in the logarithmic parameterization using the Planck data as described above, and imposing the Maximum-Entropy prior described in \ref{sec:methodology_priors} for the derived quantities $|u|_{\textnormal{max}}$ and $|s|_{\textnormal{max}}$. 

\begin{figure*}[ht]
    \centering
    \includegraphics[width=0.8\textwidth]{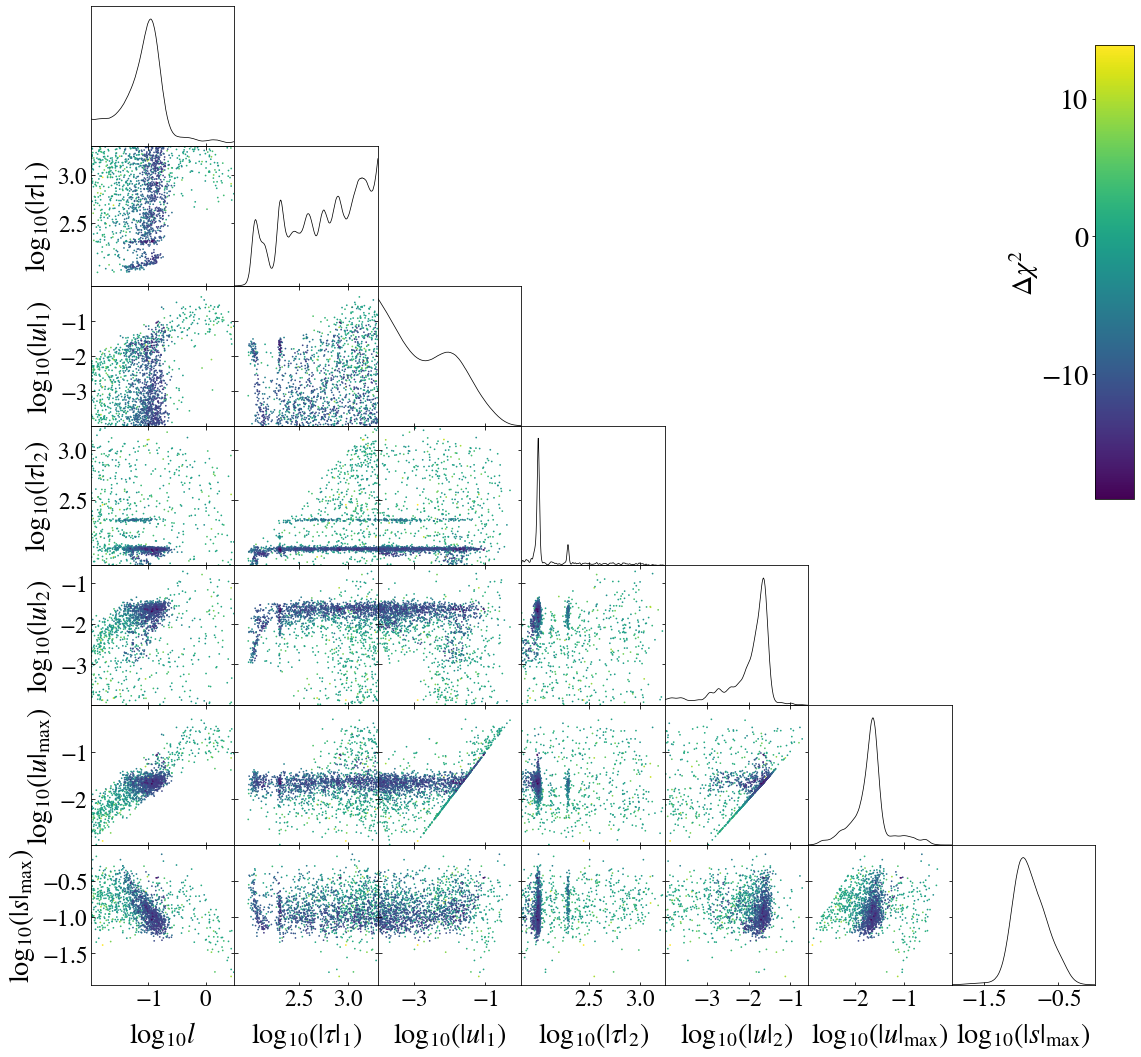}
    \caption{Posterior distribution for the reconstruction of the speed of sound's profile $u(\tau)$ using 2 training nodes and the logarithmic parametrization shown in Fig.\ \ref{fig:pipeline}. We use $\Delta \chi ^2 = \chi^2_\textnormal{model}-\chi^2_\textnormal{baseline}$ as the variable for the scatter plot's colour scale, the reference $\chi^2_\textnormal{baseline}$ corresponding to the MAP of the baseline $\Lambda$CDM model to the same datasets. We show the parameters of the training nodes $(\tau_i, u_i)$ and the correlation length $l$ (described in \ref{sec:methodology_c_s}; priors in \ref{sec:methodology_priors}). We also show the posteriors of the EFT parameters $(|u|_{\text{max}}, |s|_{\text{max}})$ (described in section \ref{sec:theory}, and not sampled directly, but derived from the nodes parameters). It can be seen how longer correlation lengths (broader reductions) lead to lower values of $|s|_{\text{max}}$, and vice-versa. The posterior distributions for different numbers of nodes display similar patterns. For all cases, the posterior distributions are clearly multi-modal.}
    \label{fig:2nodes_triangle}
\end{figure*}{}

When presenting our results, we use an effective $\Delta \chi ^2$ where we have subtracted the $\chi^2$ of the MAP of the featureless $\Lambda$CDM (obtained by using a minimizer) for the same likelihood combination (see Sec.\ \ref{sec:methodology_codes}). Note that this effective $\Delta \chi ^2$ is not meant for model selection purposes and it is used for illustration only. As an example, a triangle plot of the posterior distribution for the 2-nodes case can be seen in Fig.\ \ref{fig:2nodes_triangle}.

\begin{figure*}[ht]
    \centering
    \includegraphics[width=0.9\textwidth]{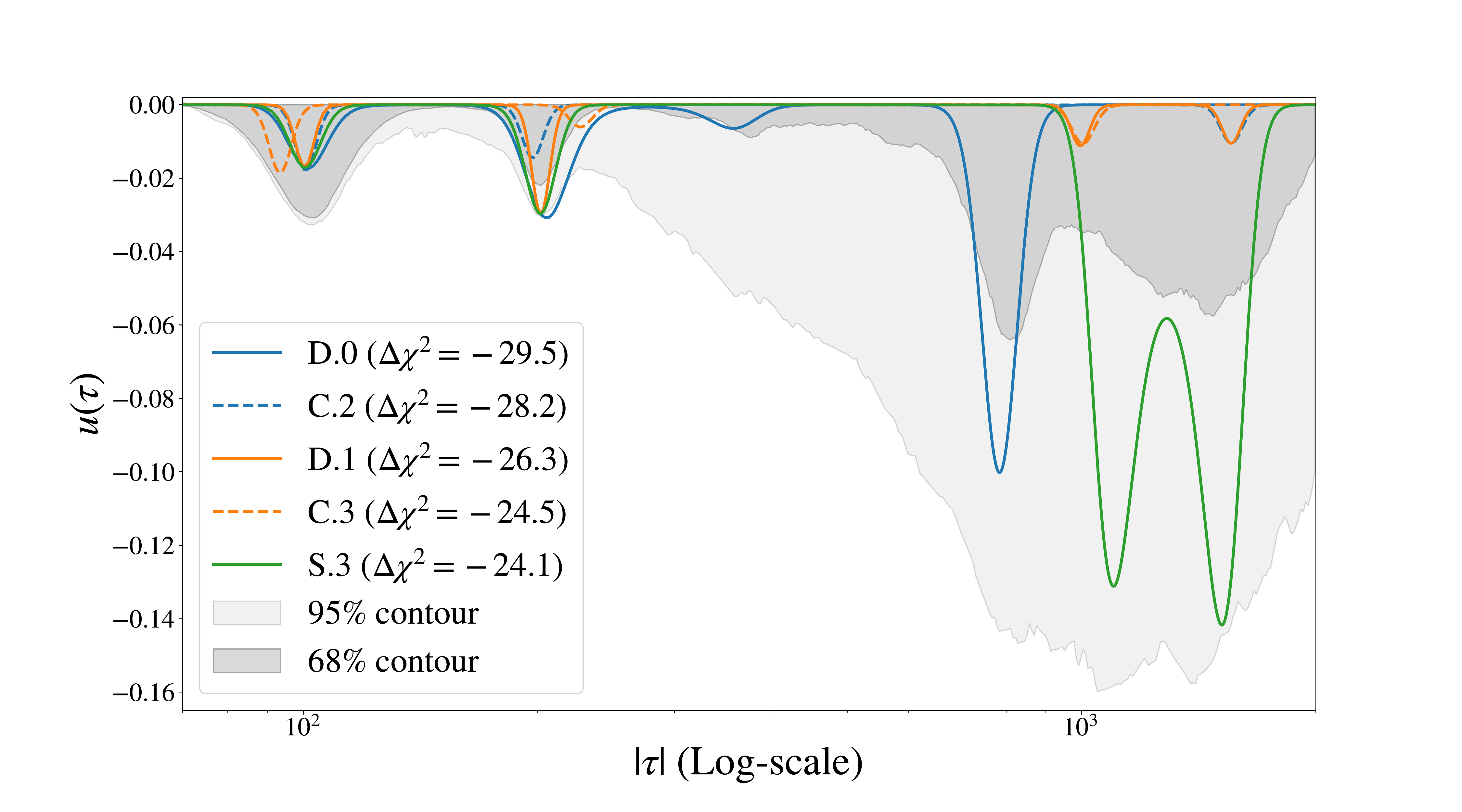}
    \caption{Reconstruction of the inflaton's speed of sound profile $u(\tau)$ based a 4-nodes GP, where the confidence contours ($68\%$ and $95\%$) are shown. We are able to constrain the shape the inflaton's speed of sound more stringently at late times (up to $\sim-200$), whereas the confidence intervals get larger at earlier times (i.e: starting from $-800$). This difference in the constraining power between early and late conformal times is mostly due to early-time reductions being associated to low-multipole features where cosmic variance is largest. The best \textit{maxima a posteriori} are also plotted on top of the confidence contours: two 3-dipped cases (labelled C.2, dashed blue, and C.3, dashed orange), two 4-dipped cases (labelled D.0, solid blue, and D.1, solid orange) and a 3-dipped case in which one of the dips possesses some substructure (labelled S.3, solid green). For additional fits and a more detailed presentation of them see appendix \ref{sec:appendix_A}.}
    \label{fig:reconstruction}
\end{figure*}{}

We have reconstructed the inflaton's speed of sound profile $u(\tau)$ using up to four training nodes. We have stopped there after checking that the Akaike Information Criterion (AIC) \cite{1100705} has a minimum for three training nodes and stabilizes after that. The profile $u(\tau)$ shows different patterns depending on how many training nodes are used in the GP reconstruction. We have decided to classify all possible profiles $u(\tau)$ based on whether they show differentiated and non-overlapping reductions (that we denominate \textit{dips}) or they present some kind of substructure:

\begin{itemize}
    \item \textbf{One single dip}: usually present at either late values of conformal time ($-100$, $-200$, $-400$), or at earlier times ($-800$, $\sim -1000$). Early-time dips produce features in the CMB power spectra localised in $\ell$'s up to the first acoustic peak, whereas features from late-time dips affect the power spectra along the full $\ell$ range. Similar profiles were already found in previous studies (see Sec.\ \ref{sec:checks}). Details on this posterior modes can be found in appendix \ref{sec:1mode}.
    
    \item \textbf{Combination of non-overlapping reductions (2, 3 and 4 dips)}: appearing when more than one training node is used, they consist of consecutive, isolated reductions in the speed of sound.\footnote{Notice that the number of training nodes is not always equal to the number of dips: reconstructions with $m$ dips found with $m$ GP nodes usually re-appear as posterior modes in the $m+1$ GP nodes case, where one of the nodes is placed at $u_i \approx 0$.} Details can be found in appendices \ref{sec:2modes}, \ref{sec:3modes} and \ref{sec:4modes}. These combinations can be classified as (for details see appendix \ref{sec:appendix_A}):
    \begin{itemize}
        \item \textit{All dips at late conformal times}: when at least two training nodes are considered, there is a preference for two of the possible dips remaining at late-time values of $\tau_i$, combining either $-100$ and $-200$, or $-400$ and $-200$. Their effect in the CMB power spectrum overlap each other along a large range of $\ell$'s.
        \item \textit{Combination of early- and late-time dips:} these appear typically as a combination of features at both low $\ell$'s (from the early-time dips) and high $\ell$'s (from the late time ones), e.g.\ from the presence of dips both at $-800$ and $-100$.
    \end{itemize}
    \item \textbf{Dips with substructure:} We have found some \textit{maxima a posteriori} where the reconstructed $u(\tau)$ does not show clearly separated reductions, but a more complex profile with some degree of substructure. These substructures are presented either at early and late $\tau_i$, trying to fit some of the characteristic features of the CMB angular power spectrum (i.e: $\ell \approx 20-40$ feature). The fits are presented in Appendix A, in subsection \ref{sec:substructure}.
\end{itemize}

As noticed in previous works \cite{2017PhRvD..96h3515T, 2015MNRAS.453.4384H}, we do not have a highly predictive posterior of the maximum of the rate of change of the sound speed, $|s|_\textnormal{max}$, whose value is mostly constrained by the prior information. By contrast, the positions of the nodes (the oscillation frequency of the features in the power spectrum) are tightly constrained within each of the multiple posterior modes, specially for nodes at late conformal time.

Using the sampling results of the profile $u(\tau)$ with four training nodes in the GPs, we have reconstructed the allowed confidence contours for $u(\tau)$ given Planck 2018 data. The result can be seen in Fig.\ \ref{fig:reconstruction}. As expected from the 1-D marginalized posterior distributions, the confidence contours are narrower around $\tau_i=[-100, 200]$. These modes were found in every single reconstruction of the inflaton's speed of sound independently of the number of training nodes (and were also observed in previous studies \cite{2017PhRvD..96h3515T}), and usually show the highest individual dip $\Delta \chi ^2$ with respect to $\Lambda$CDM (since they produce features at a long range of $\ell$ for which Planck has low error bars). On the other hand, the confidence contours are broader for earlier conformal times $\tau_i < -400$. This is the range of $\tau$ where we have found the modes at $\tau_i = [-800, -1000]$ and some degree of substructure. In this range, the posterior distributions are not very predictive (see again Fig.\ \ref{fig:2nodes_triangle}, where the posterior peaks are small for $\tau_i<-800$), since they produce low-multipole features hidden by cosmic variance.

\section{Conclusions}\label{sec:conclusion}

We have searched for features in the primordial power spectrum as given by the last release of Planck 2018 data. Following an EFT of inflation approach, we have focused our search on features coming from reductions of the sound speed of the inflaton, assuming these reductions to be small, mild and transient. These feature templates were not tested by the Planck Collaboration.

We have improved over previous studies (which used a single-reduction Gaussian ansatz) by developing a reconstruction technique for the speed of sound's profile based on Gaussian Processes. We have also marginalised over the slow roll parameters to allow for a dynamical prior. In this new pipeline, the parameters of the reconstruction (the position of the training nodes and the correlation length) are fitted to the Planck 2018 data. The physical constraints of the model are imposed on the reconstruction parameters by means of a Maximum-Entropy prior defined on the EFT quantities $(|u|_{\textnormal{max}}, |s|_{\textnormal{max}})$, which define the consistency bounds of the model. We have also tracked as derived parameters $n_s$ and $r$.

This template-free reconstruction of $u(\tau)$ has allowed us to make an exhaustive search of more flexible features' templates, constrained only by EFT conditions. The analysis of the result of Bayesian parameter inference on the Planck 2018 data has demonstrated that there are many possible different and complex $u(\tau)$ profiles which are consistent with Planck's CMB power spectra. As expected, none of these fits is preferred with respect to $\Lambda$CDM (their $\Delta \chi ^2$'s are not significant), although show some interesting results in terms of new feature templates. First, we have argued that there is a strong preference for two consecutive reductions of the speed of sound to coexist at late times around $\tau_i \approx -200$ and $\tau_i \approx -100$. Also, combinations of modes at late conformal time $\tau_i \approx -100$ and early conformal time $\tau_i \approx -800$ are also possible. Second, we have found certain profiles which show some degree of sub-structure at early and late conformal times. Finally, we have been able to obtain reconstruction confidence contours for the $u(\tau)$ profile given the results obtained with four training nodes.

In the future, we plan to exploit this robust and novel pipeline in the search of features using new sets of data (in particular, Large Scale Structure surveys or the CMB bispectrum). Furthermore, the improvement of current data (for example, the polarization of the CMB) will also help to reduce the noise and, therefore, the uncertainty we have at large scales. If the noise is reduced, we could discern how realistic the reductions at earlier conformal times are. Moreover, we also consider introducing new features coming from a variable first slow-roll parameter \cite{epsilon} to perform a joint search of both patterns: features induced by a variable $c_s(\tau)$ and $\epsilon(\tau)$.

%%%%%%%%%%%%%%%%%%%%%%%%%%%%%%%%%%%%%%%%%%%%%%%%%%%%%%%%%%%%%%%%%%%%%%%%%%
\begin{acknowledgments}
%%%%%%%%%%%%%%%%%%%%%%%%%%%%%%%%%%%%%%%%%%%%%%%%%%%%%%%%%%%%%%%%%%%%%%%%%%
We thank Dhiraj Hazra for insightful discussion over the likelihoods nuisance parameters. We thank Juan Claramunt Gonzalez for useful comments concerning the mathematical notation used in the methodology. We acknowledge Santander Supercomputacion support group at the University of Cantabria who provided access to the supercomputer Altamira Supercomputer at the Institute of Physics of Cantabria (IFCA-CSIC), member of the Spanish Supercomputing Network, for performing simulations/analyses. GCH acknowledges support from the Delta Institute for Theoretical Physics (D-ITP consortium), a program of the Netherlands Organization for Scientific Research (NWO, OWC). JT acknowledges partial support from the European Research Council under the European Union's Seventh Framework Programme (FP/2007-2013) / ERC Grant Agreement No.\ [616170]. AA is partially supported by the Netherlands' Organization for Scientific Research (NWO/FOM), by the Basque Government/Eusko Jaurlaritza (IT- 979-16), and by the Spanish Ministry MINECO FPA2015-64041-C2-1P (MINECO/FEDER) and PGC2018-094626-B-C21 (MCIU/AEl/FEDER,UE).
\end{acknowledgments}

%%%%%%%%%%%%%%%%%%%%%%%%%%%%%%%%%%%%%%%%%%%%%%%%%%%%%%%%%%
\appendix
\section{detailed results up to 4 nodes of $u(\tau)$ in log-log parameterization}\label{sec:appendix_A}

In this appendix, we explain in detail the several \textit{maxima a posteriori} found during the sampling runs when the profile of the inflaton's speed of sound $u(\tau)$ was reconstructed using Gaussian Processes up to 4 training nodes.

\begin{figure*}[]
    \centering
    \includegraphics[width=0.65\textwidth]{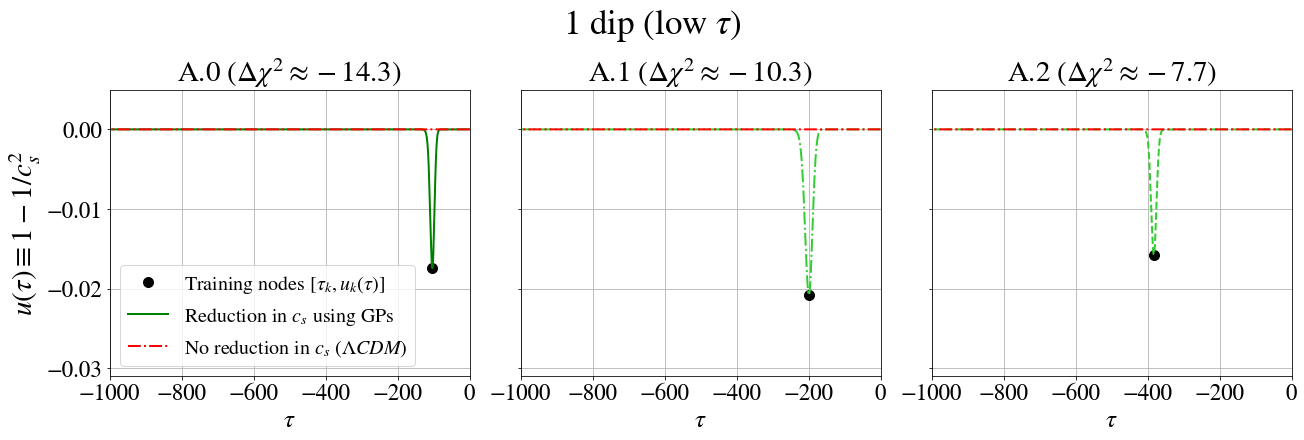}
    \includegraphics[width=0.65\textwidth]{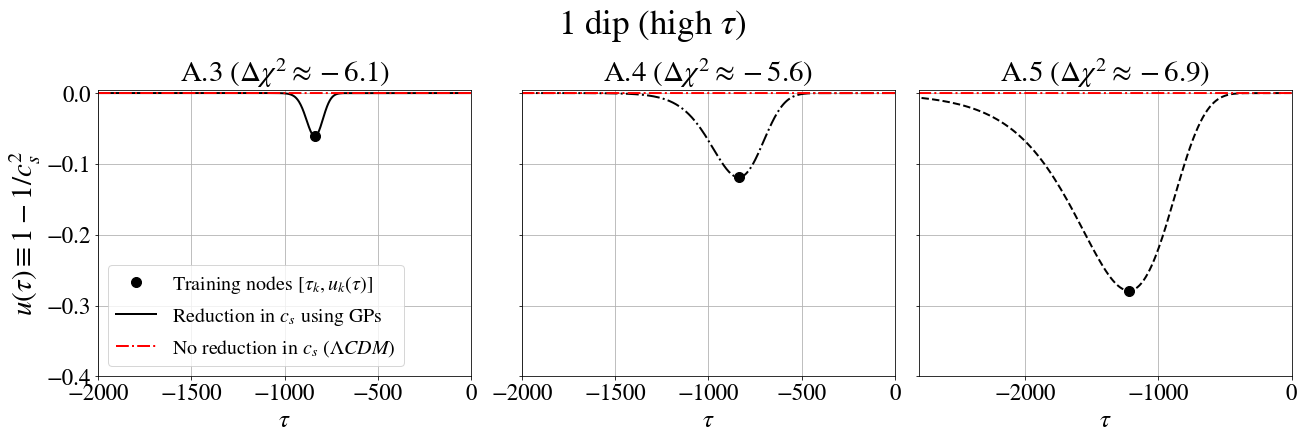}
    \includegraphics[width=0.7\textwidth]{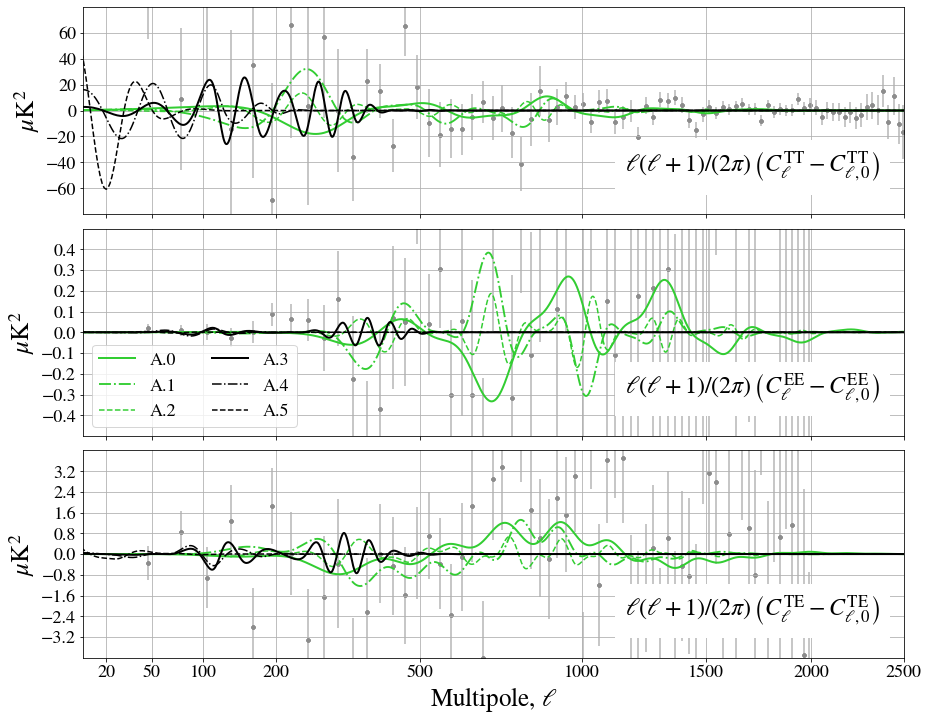}
    \caption{One single dip, one training node. \textbf{Top:}  Different profiles $u(\tau)$ for the 6 \textit{maxima a posteriori} when only 1 training node is used (and consequently only one dip is visible). The reconstruction is done following the logarithmic parametrization explained in Sec. \ref{sec:methodology_c_s}. We found a principal MAP and 5 other fits when the multimodal posterior distribution is further analysed (see, for example, Fig.\ \ref{fig:2nodes_triangle}, where other peaks in the posterior distribution are visible). \textbf{Bottom}: Differences in the CMB temperature (TT), E-polarization (EE) and cross-correlated power spectra (TE) between the MAP to the Planck 2018 data and the featureless $\Lambda$CDM baseline model for the reconstructed speed of sound profiles $u(\tau)$ A.0 - A.5 shown above. Notice how these profiles fit small deviations from $\Lambda$CDM at low and high multipoles $\ell$. The same color and line-style correspondence between the $u(\tau)$ profiles and the differences in the CMB spectra has been used.}
    \label{fig:1modes}
\end{figure*}{}

\begin{figure*}[htp!]
    \centering
    \includegraphics[width=0.8\textwidth]{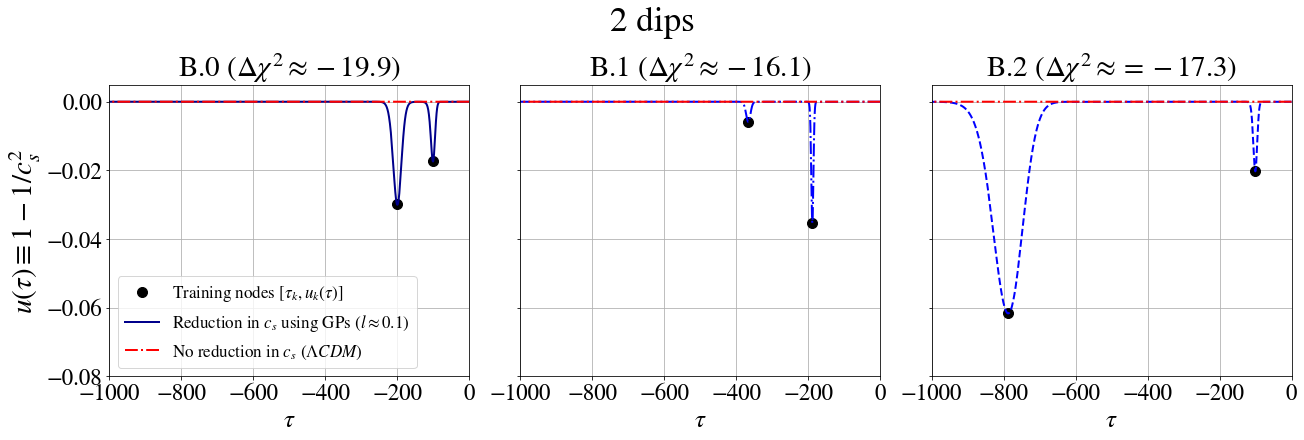}
    \includegraphics[width=0.8\textwidth]{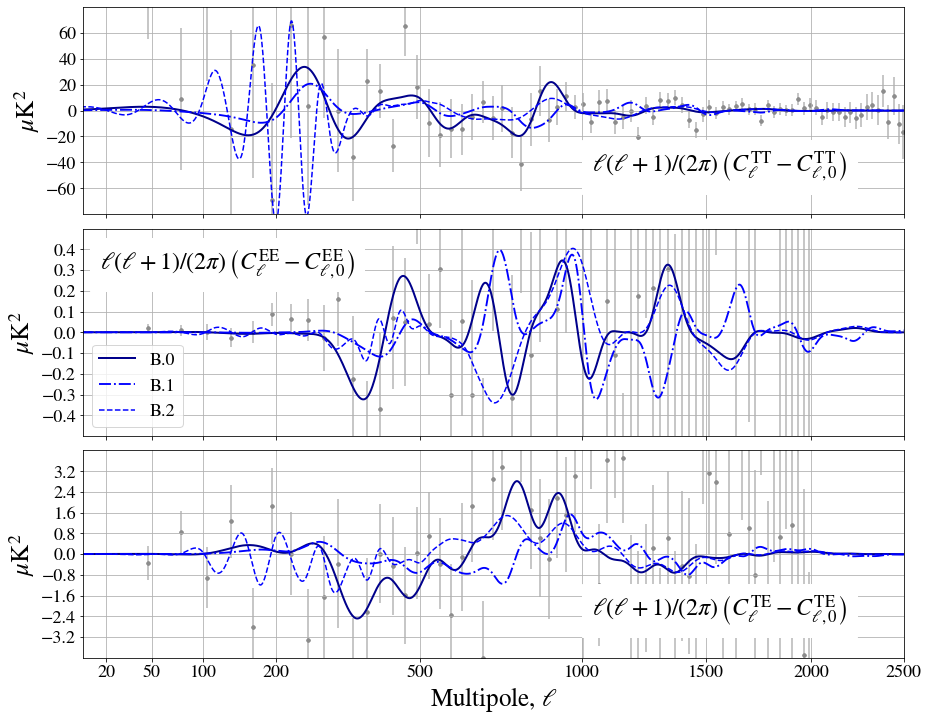}
    \caption{2 non-overlapping dips, 2 training nodes. \textbf{Top:}  Different profiles $u(\tau)$ for the 3 \textit{maxima a posteriori} when only 2 training nodes are used and only two clearly different dips are observed. The reconstruction is done following the logarithmic parametrization explained in Sec. \ref{sec:methodology_c_s}. We found a principal best fit and 2 other fits when the multimodal posterior distribution is further studied (see, for example, Fig.\ \ref{fig:2nodes_triangle}, where other peaks in the posterior distribution are visible). \textbf{Bottom}: Differences in the CMB temperature (TT), E-polarization (EE) and cross-correlated power spectra (TE) between the best fit to the Planck 2018 data and the featureless $\Lambda$CDM baseline model for the reconstructed speed's of sound profile $u(\tau)$ shown above. Notice how these profiles fit small deviations from $\Lambda$CDM at low and high multipoles $\ell$. The same color and line-style correspondence between the $u(\tau)$ profiles and the differences in the CMB spectra has been used.}
    \label{fig:2modes}
\end{figure*}{}

\begin{figure*}[htp!]
    \centering
    \includegraphics[width=0.8\textwidth]{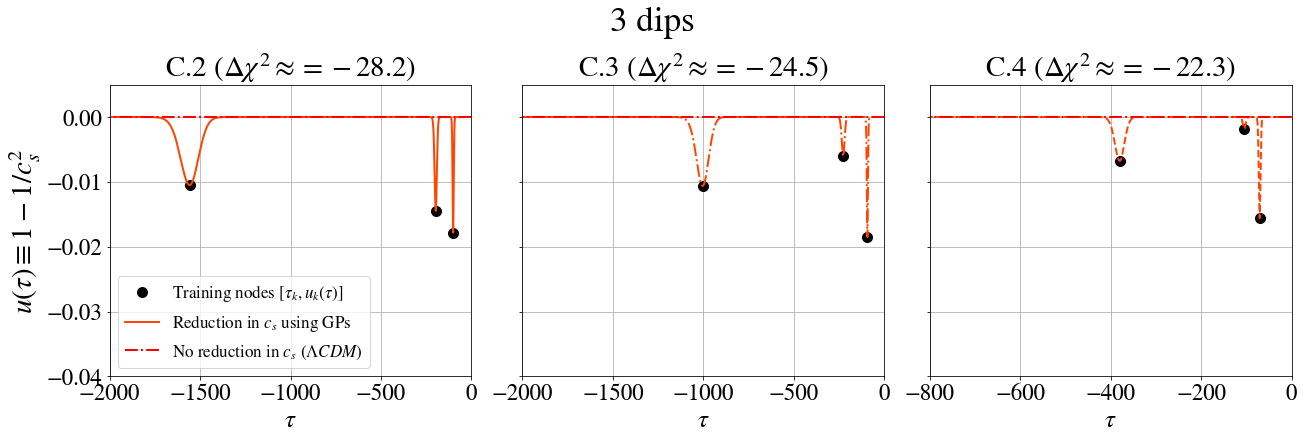}
    \includegraphics[width=0.8\textwidth]{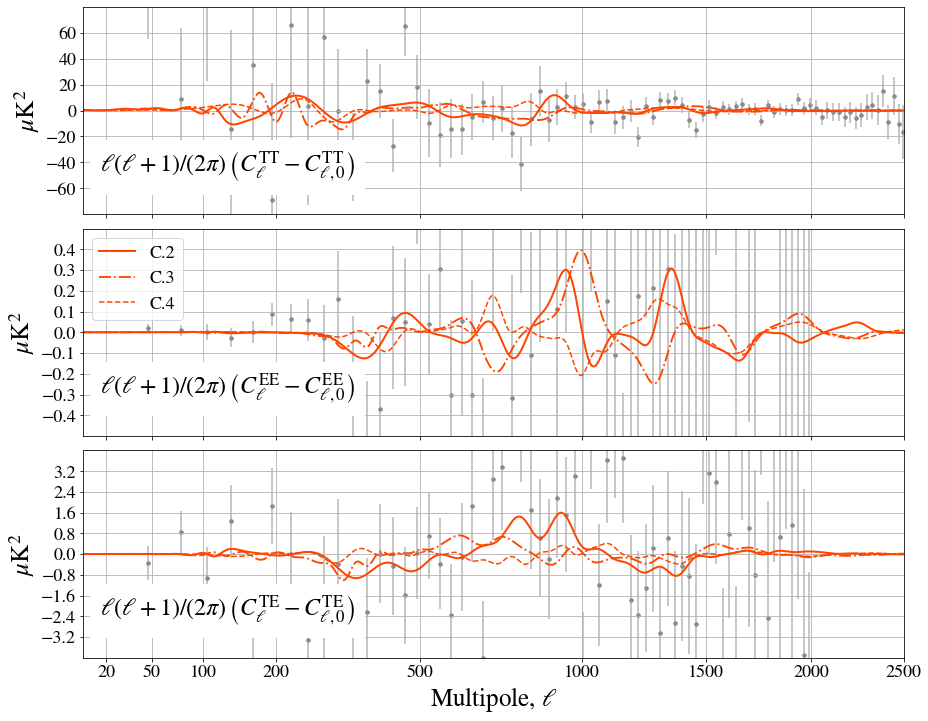}
    \caption{3 non-overlapping dips, 3 training nodes. \textbf{Top:}  Different profiles $u(\tau)$ for the 3 \textit{maxima a posteriori} when only 3 training nodes are used and three differentiated dips are observed. The reconstruction is done following the logarithmic parametrization explained in Sec. \ref{sec:methodology_c_s}. We found 3 MAP when the corresponding multimodal posterior distribution is further studied (see, for example, Fig.\ \ref{fig:2nodes_triangle}, where other peaks in the posterior distribution are visible). \textbf{Bottom}: Differences in the CMB temperature (TT), E-polarization (EE) and cross-correlated power spectra (TE) between the best fit to the Planck 2018 data and the featureless $\Lambda$CDM baseline model for the reconstructed speed's of sound profile $u(\tau)$ shown above. Notice how these profiles fit small deviations from $\Lambda$CDM at low and high multipoles $\ell$. The same color and line-style correspondence between the $u(\tau)$ profiles and the differences in the CMB spectra has been used.}
    \label{fig:3modes}
\end{figure*}{}

\begin{figure*}[htp!]
    \centering
    \includegraphics[width=0.8\textwidth]{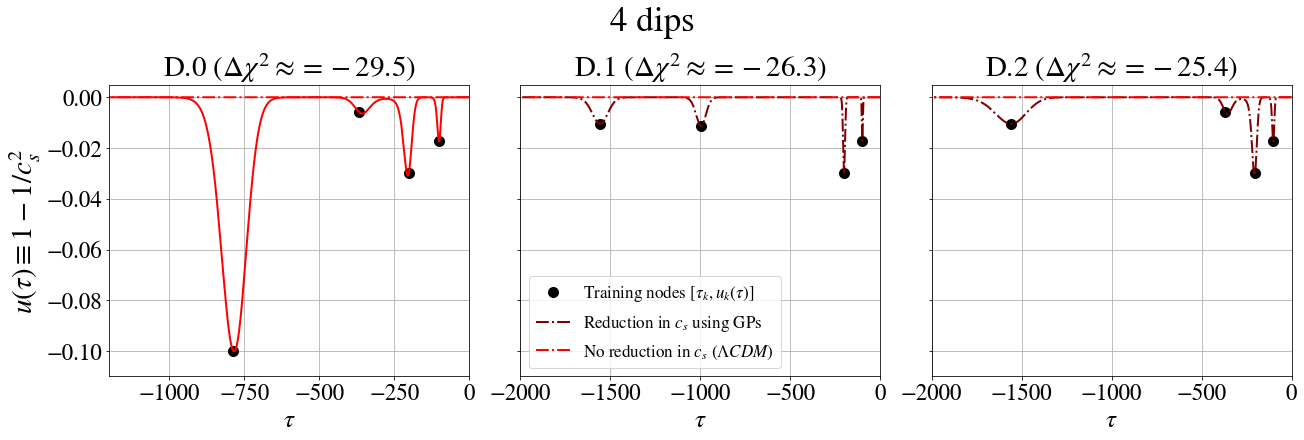}
    \includegraphics[width=0.8\textwidth]{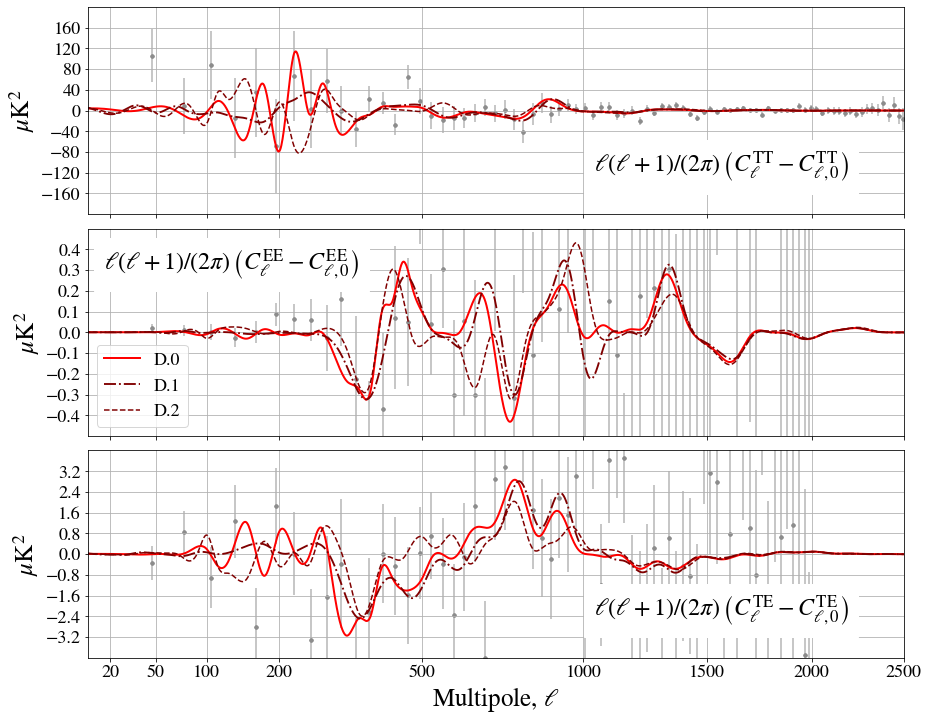}
    \caption{4 non-overlapping dips, 4 training nodes. \textbf{Top:}  Different profiles $u(\tau)$ for the 3 \textit{maxima a posteriori} when only 4 training nodes are used and four differentiated dips are observed. The reconstruction is done following the logarithmic parametrization explained in Sec. \ref{sec:methodology_c_s}. In this case, we observe how possible reductions at $\tau \approx -100, -200, -400$ and $-800$ can consecutively take place. \textbf{Bottom}: Differences in the CMB temperature (TT), E-polarization (EE) and cross-correlated power spectra (TE) between the best fit to the Planck 2018 data and the featureless $\Lambda$CDM baseline model for the reconstructed speed's of sound profile $u(\tau)$ shown above. Notice how these profiles fit small deviations from $\Lambda$CDM at low and high multipoles $\ell$. The same color and line-style correspondence between the $u(\tau)$ profiles and the differences in the CMB spectra has been used.}
    \label{fig:4modes}
\end{figure*}{}

\begin{figure*}[htp!]
    \centering
    \includegraphics[width=0.8\textwidth]{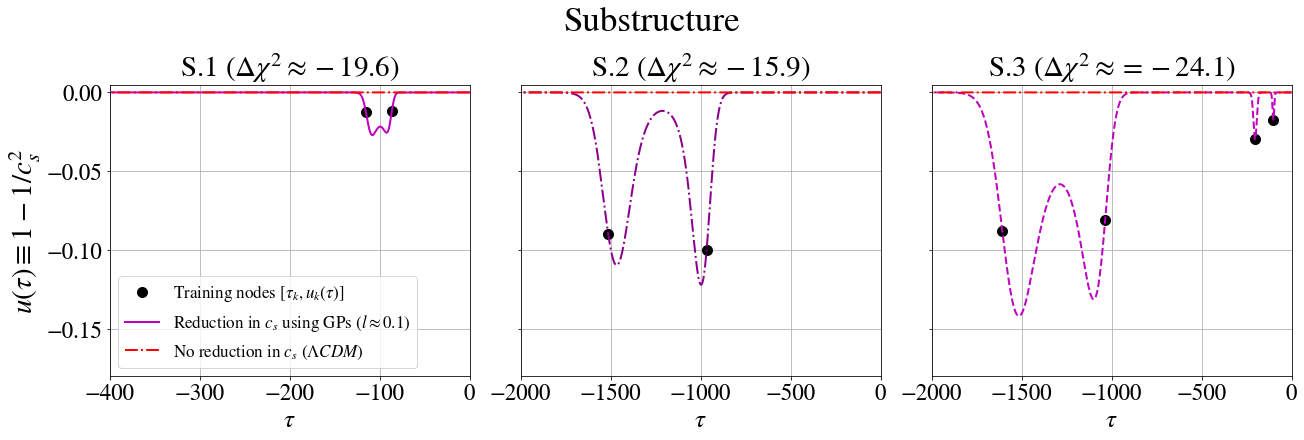}
    \includegraphics[width=0.8\textwidth]{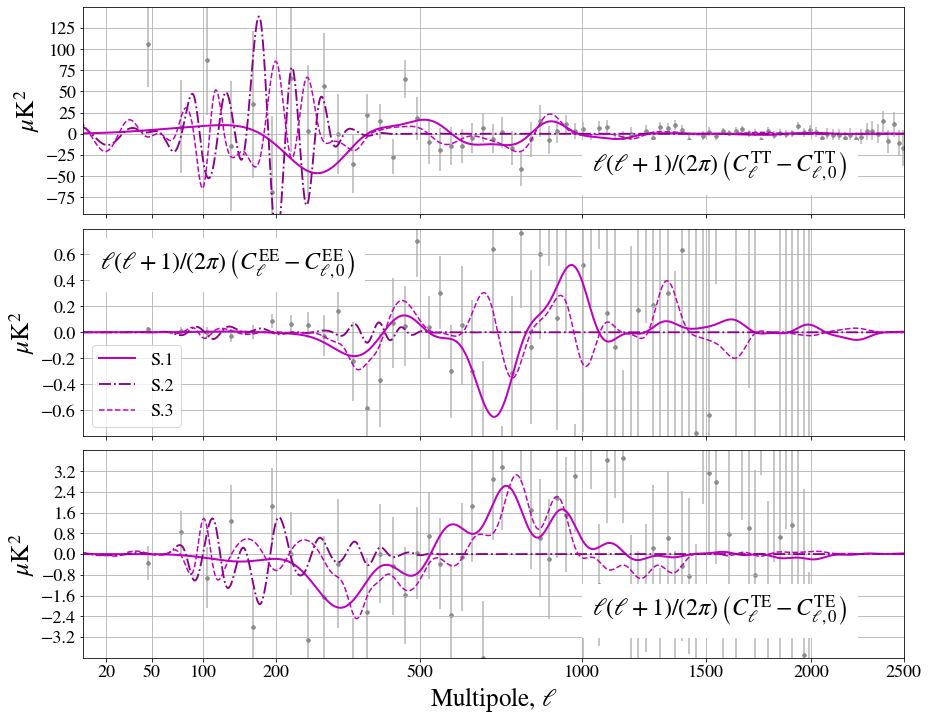}
    \caption{Profiles with substructure (2 training nodes and 4 training nodes). \textbf{Top:} Different profiles $u(\tau)$ for 2 \textit{maxima a posteriori} when only 2 or training nodes are used and the profiles of $u(\tau)$ show some grade of substructure. The reconstruction is done following the logarithmic parametrization explained in Sec. \ref{sec:methodology_c_s}. We found 2 fits (one an earlier conformal time and a another one at late conformal time), when the corresponding multimodal posterior distributiosn are further studied (see, for example, Fig.\ \ref{fig:2nodes_triangle}, where other peaks in the posterior distribution are visible). \textbf{Bottom}: Differences in the CMB temperature (TT), E-polarization (EE) and cross-correlated power spectra (TE) between the best fit to the Planck 2018 data and the featureless $\Lambda$CDM baseline model for the reconstructed speed's of sound profile $u(\tau)$ shown above. Notice how these profiles fit small deviations from $\Lambda$CDM at low and high multipoles $\ell$. The same color and line-style correspondence between the $u(\tau)$ profiles and the differences in the CMB spectra has been used.}
    \label{fig:substructure}
\end{figure*}{}

\subsection{One dip (\textit{denoted by “A”})}\label{sec:1mode}
These profiles of $u(\tau)$ show only one single reduction of the inflaton's speed of sound or \textit{one dip}. These reductions can be found using just one training node in the Gaussian Process. The modes at late conformal time $\tau_i$ (see the first row of Fig.\ \ref{fig:1modes}) present a well defined oscillation frequency $\tau_i$ (at -100, -200 and -400). The value of $|u|_\textnormal{max}$ is around 0.02 and the rate of change in the speed of sound $|s|_\textnormal{max} \gg |u|_\textnormal{max}$. These dips are exactly reproduced with the linear parametrization of the reconstruction of $u(\tau)$. They were already listed during previous searches using Planck 2013 and 2015 data \cite{2015PhRvD..91f4039H, 2017PhRvD..96h3515T}. In particular, the mode corresponding at $\tau_i \approx -400$ was identified faintly in \cite{2017PhRvD..96h3515T}. These modes are present in a broad multipole $\ell$ range ($\ell \approx [100-2000])$, fitting some structures in the temperature and polarization data.

Modes at early conformal time $\tau_i$ (-800, -1000) (see the second row of Fig.\ \ref{fig:1modes}) are found, but are more poorly constrained and with worse $\Delta \chi ^2$ with respect to $\Lambda$CDM. Mode A.3 at $\tau_i \approx -800$ shows similar characteristics to the modes at a late conformal time (small amplitude and same behaviour in the EFT parameter $s$). It fits an apparent oscillating structure of the temperature $C_\ell$ at the first acoustic peak. This mode is also found with the linear parametrization and in previous searches in \cite{2015PhRvD..91f4039H, 2017PhRvD..96h3515T}. The modes at -800 and -1000, with a larger amplitude, have $|s|_\textnormal{max} \approx |u|_\textnormal{max}$. They slightly differ from the modes found in previous studies. The main reason is that $u(\tau)$ reconstructed using the logarithmic parametrization differs from the linear one at high values of $\tau_i$. The modes at -1000 try to fit the characteristic $\ell \approx 20-40$ structure of the CMB temperature angular power spectrum. As identified in \cite{2017PhRvD..96h3515T}, this kind of features impose a tighter upper limit on the scalar-to-tensor ratio $r$ (see the tables on this section), although these are still within the analogous bounds in $\Lambda$CDM.

\subsection{Two dips (\textit{denoted by “B”})}\label{sec:2modes}

In this case, the inflaton would suffer two consecutive reductions of the speed of sound (due to, for instance, two consecutive turns in the field space). In the previous study using Planck 2015 \cite{2017PhRvD..96h3515T}, it was pointed out that, a priori, the features due to single reductions of the sound speed that do not overlap can in principle co-exist. These combinations would be modes at early $\tau_i$ with another late mode (i.e: -1000 and -100). However, the results of the reconstruction using two training nodes show a richer picture (see Fig.\ \ref{fig:2modes}). We have identified that the modes at -100 and -200 can result from an overlapping feature that is preferred by the data, and thus, it is the overall MAP for the 2-nodes reconstruction. The dips at -200 and -400 can also co-exist (mode B.1), with a worse $\Delta \chi ^2$. These two overlapping features fit TT, TE and EE structure across a large range of $\ell$. On the other hand, there is a possible combination of the modes at -800 and -100 (see the first row of Fig.\ \ref{fig:2modes}). This $u(\tau)$ profile includes the fitting of the apparent oscillations around the first acoustic peak and small deviations across the rest of the multipole scale. All of these combinations of modes fulfill $|s|_\textnormal{max} \gg |u|_\textnormal{max}$. 

\subsection{Three dips (\textit{denoted by “C”})}\label{sec:3modes}

When the profile of the inflaton's reduction of the speed of sound $u(\tau)$ is reconstructed using 3 training nodes, we find more complex profiles. The usual dips at late $\tau_i$ around -100 and -200 combine to mild and small reductions at early conformal times around -1000 and -1500 (see modes C.2 and C.3 of Fig.\ \ref{sec:3modes}, respectively). The combinations that are preferred by the data are those whose earlier training node $\tau_i$ is placed around $u_i\approx -0.01$. These small dips at early $\tau_i$ were not found alone during the search using one training node (modes A shown in Fig.\ \ref{fig:1modes}). However, these modes at early conformal times are loosely constrained and the confidence intervals are large. Overall, these profiles fits the data very similarly as the single standing modes A.0 and A.1, as well as the combination B.0. Furthermore, we have also verified that the modes A.0 and A.2 can co-exist (contrary to the case of reconstruction with 2 training nodes, where the combination -100 and -400 was not found) if a very small mode is added close to -100. All these profiles presented in Fig.\ \ref{fig:3modes} are in the limit $|s|_\textnormal{max} \gg |u|_\textnormal{max}$. Finally, the 2-dip profiles explained above are also reproduced when we run with 3 training nodes.

\subsection{Four dips (\textit{denoted by “D”})}\label{sec:4modes}

We find similar profiles for $u(\tau)$ as the ones for 3 training nodes, adding one extra dip and finding the remaining possible combination of nodes at late $\tau_i$ around -100, -200 and -400 with modes at earlier times at -800, -1000 or -1500 (see modes D.0, D.1 and D.3 of Fig.\ \ref{fig:4modes}). Thus, we have corroborated that the modes A.0, A.1, A.2 and A.4 can, in principle, co-exist. Still, the small modes at early $\tau_i\approx -1000, -1500$ are less likely to show up in the posterior, as they are poorly constrained given the data. It is worth mentioning that we can mostly reproduce all the different profiles found during the search using up to three nodes when four training nodes are used. In this case, one, two or three training nodes are placed in such a way the corresponding profile looks very similar to the cases A, B or C (either the training node is placed close to $u_i \approx 0$ or close to the previous training mode itself).\\

\subsection{Reductions with substructure (\textit{denoted by “S”})}\label{sec:substructure}

Apart from concatenations of transient reductions in the speed of sound, we have also observed some possible fits which show more complicated feature patterns according to the data. These are profiles of $u(\tau)$ that do not clearly show full dips but have some kind of substructure (see the upper row of Fig.\ \ref{fig:substructure}, profile S.1). We find a sub-structured maxima a posteriori at a late conformal time (centred around $\tau_i \approx -100$), which resembles the mode A.0 but with two small sub-reductions. Similarly to A.0, the limit of the EFT functions is $|s|_\textnormal{max} \gg |u|_\textnormal{max}$.

Motivated by the loose constraints of the training nodes in the range of early conformal times $-800<\tau_i<-3000$, we have launched a GPs reconstruction using two training nodes, which are restricted to remain in the range  $-3500 < \tau_i<-990$, obtaining the profile S.2 of Fig.\ \ref{fig:substructure}. This profile have the particularity that $|s|_\textnormal{max} \ll |u|_\textnormal{max}$. This profile tries to fit not only the structure of the CMB TT angular power spectrum around $\ell \approx 20-40$ but also the apparent structure of the first acoustic peak in the TT and TE data. To answer the question if it is possible the substructure mode S.2 to co-exist with any of the modes at late conformal time $\tau_i$, we have relaunched the GPs reconstruction with 4 training nodes, constraining the two early ones in the range  $-3500< \tau_1< \tau_2 <-990$. When this constraint is imposed, the later training nodes $\tau_3$ and $\tau_4$ are placed clearly around -100 and -200. The resulting profile (see mode S.3 in Fig.\ \ref{fig:substructure}) can fit the CMB data at low $\ell$ but also in a broader range similarly to the case of the mode A.0, increasing the statistical significance $\Delta \chi ^2$. In this case, the EFT limit is reverted to $|s|_\textnormal{max} \gg |u|_\textnormal{max}$ due to the narrow mode at -100.

\begin{table*}[h]
\begin{tabular}{|r|ccc|ccc|ccc|ccc|}
\hline
\multicolumn{1}{|c|}{Name} &  \multicolumn{3}{c|}{\textbf{A.0 ($\Delta \chi ^2 = -14.3$)}} & \multicolumn{3}{c}{\textbf{B.0 ($\Delta \chi ^2 = -19.9 $)}} & \multicolumn{3}{|c|}{\textbf{C.2 ($\Delta \chi ^2 = -28.2$)}} & \multicolumn{3}{|c|}{\textbf{D.0 ($\Delta \chi ^2 = -29.5$)}} \\
\hline
\hline
\multicolumn{1}{|c|}{Parameters} & Center      & lower     & upper     & Center      & lower      & upper     & Center      & lower      & upper  & Center      & lower      & upper    \\
\hline
$\log_{10}l$               & -0.972      & -1.997     & 0.300      & -0.995      & -1.989      & 0.282      &    -1.231 & -1.211 & -0.754  &    -0.994 & -1.165 & -0.917    \\
$\log_{10}(|\tau_1|)$            & 2.014       & 1.959      & 2.299      & 2.305       & 1.993  & 3.299     &     3.192  & 2.834  & 3.297  & 2.896 &  2.397  &  3.100    \\
$\log_{10}(|u_1|)$             & -1.760      & -2.687     & -0.271     & -1.528      & -3.999      & -0.593   &    -1.979 & -1.952 & -0.851 & -1.101 & -3.997 & -0.597   \\
$\log_{10}(|\tau_2|)$            & -           & -          & -          & 2.002       & 1.845       & 3.145&    2.295  & 2.237  & 2.395  & 2.564  & 2.285 &  3.301  \\
$\log_{10}(|u_2|)$              & -           & -          & -          & -1.764      & -3.843      & -0.977 &     -1.840 & -3.958 & -1.495 &  -2.229 & -3.987 & -0.404   \\
$\log_{10}(|\tau_3|)$           & -           & -          & -          & -           & -           & -     &    2.003  & 1.963  & 2.030  & 2.564  &  2.492 &  2.629     \\
$\log_{10}(|u_3|)$           & -           & -          & -          & -           & -           & -          &   -1.749 & -2.039 & -1.495  & -2.229 & -1.854 & -0.741   \\
$\log_{10}(|\tau_4|)$           & -           & -          & -          & -           & -           & -     &    - & - & - &  2.002       & 1.983      & 3.310   \\
$\log_{10}(|u_4|)$           & -           & -          & -          & -           & -           & -          &   - & - & - & -1.764      & -2.693     & -0.256    \\
$\log_{10}(|u|_\mathrm{max})$                & -1.462      & -2.694     & -0.282     & -1.523      & -2.587      & -0.500     & -1.621 & -1.810 & -0.851  & -0.998 & -1.831 & -0.804   \\
$\log_{10}(|s|_\mathrm{max})$                & -0.768      & -1.442     & -0.235     & -0.813      & -1.352      & -0.312     & -0.928 & -1.084 & -0.206 &  -0.266 & -1.731 & -0.471    \\
$\epsilon_1$                 & 0.000       & 0.000      & 0.007      & 0.000       & 0.000       & 0.006      &  0.001  & 0.000  & 0.005  &  0.001  & 0.000  & 0.005    \\
$\epsilon_2$                   & 0.036       & 0.021      & 0.039      & 0.036       & 0.025       & 0.040      & 0.035  & 0.027  & 0.037  & 0.032  & 0.025  & 0.038    \\
$n_s$                   & 0.964       & 0.960      & 0.968      & 0.963       & 0.959       & 0.968      & 0.963  & 0.961  & 0.967 & 0.963  & 0.963  & 0.968   \\
$r$                    & 0.002       & 0.002      & 0.111      & 0.005       & 0.002       & 0.094      & 0.015  & 0.002  & 0.078 & 0.014  & 0.002  & 0.069 \\
\hline
\end{tabular}
\caption{\textit{Maxima a posteriori} values and 68\% confidence intervals of the feature and primordial parameters for the cases where $|\Delta \chi ^2|$ is the largest, when a minimizer method is used. The correspond to 1, 2, 3 and 4 training nodes used in the GPs reconstruction, respectively.}
\end{table*}

\begin{table*}
\begin{tabular}{|r|ccc|ccc|}
\hline
\multicolumn{1}{|c|}{Number of nodes $i$} & \multicolumn{6}{|c|}{\textbf{1 training node} (late conformal time)} \\
\hline         
\hline
\multicolumn{1}{|c|}{Modes}  & \multicolumn{3}{c|}{A.1  ($\Delta \chi ^2 = -10.3$)}   & \multicolumn{3}{c|}{A.2  ($\Delta \chi ^2 = -7.7$)}           \\
\hline
 \multicolumn{1}{|c|}{Parameters}  & Center           & lower      & upper  & Center           & lower      & upper   \\
\hline
$\log_{10}l$     & -1.012       & -1.942      & -0.762     & -1.445       & -1.996      & -0.983     \\
$\log_{10}(|\tau_1|)$  & 2.299        & 2.202      & 2.300 & 2.583	& 2.501 &	2.639      \\
$\log_{10}(|u_1|)$      & -1.682       & -2.304     & -1.410 & -1.804	& -2.168	& -0.870      \\
$\log_{10}(|u|_\mathrm{max})$     & -1.592       & -2.313      & -1.412 & -1.814 &	-2.777 &	-0.282     \\
$\log_{10}(|s|_\mathrm{max})$     & -0.727       & -0.999      & -0.275 & -0.617 &	-1.442	& -0.304     \\
$\epsilon_1$     & 0.000        & 0.000       & 0.004      & 0.0001	& 0.0001 &	0.0043 \\
$\epsilon_2$      & 0.036        & 0.027       & 0.039     & 0.035 &	0.026 &	0.038  \\
$n_s$        & 0.964        & 0.961       & 0.967    & 0.965 &	0.961 &	0.968   \\
$r$         & 0.003        & 0.002       & 0.058   & 0.002 &	0.002 &	0.068   \\
\hline
\end{tabular}
\caption{1 node secondary \textit{maxima a posteriori} values, at low conformal time, and 68\% confidence intervals of the feature and primordial parameters when a minimizer method is used.}
\end{table*}

\begin{table*}
\begin{tabular}{|r|ccc|ccc|ccc|}
\hline
\multicolumn{1}{|c|}{Number of nodes $i$}          & \multicolumn{9}{|c|}{\textbf{1 training node} (early conformal time)} \\
\hline         
\hline
\multicolumn{1}{|c|}{Modes}  &  \multicolumn{3}{c|}{A.3  ($\Delta \chi ^2 = -6.1$)} & \multicolumn{3}{c|}{A.4  ($\Delta \chi ^2 = -5.6$)} & \multicolumn{3}{c|}{A.5  ($\Delta \chi ^2 = -6.9$)}          \\
\hline
 \multicolumn{1}{|c|}{Parameters}  & Center           & lower      & upper  & Center           & lower      & upper  & Center           & lower      & upper  \\
\hline
$\log_{10}l$     & -0.988       & -1.642      & -0.661 &  -0.488	&-1.066	&0.050  & -0.233 &	-1.141	&0.328         \\
$\log_{10}(|\tau_1|)$  & 2.923        & 2.402       & 3.100  & 2.931&	2.804&	2.995 &  3.087 &	3.001 &	3.299    \\
$\log_{10}(|u_1|)$      & -1.215       & -2.976 & -0.807  & -1.210&	-1.470&	-0.472 &  -0.554 &	-1.457	& -0.270          \\
$\log_{10}(|u|_\mathrm{max})$     & -1.169  & -1.880 & -0.797  & -1.367&	-1.470&	-0.474 &  -0.554 &	-1.459 & -0.282    \\
$\log_{10}(|s|_\mathrm{max})$     & -0.473  & -1.231 & -0.340  &-0.702&	-1.371&	-0.406 &  -0.614 &	-1.442 &	-0.329   \\
$\epsilon_1$     & 0.000        & 0.000       & 0.005  &0.0001	&0.0001	&0.0035 & 0.0001 & 0.0001 & 0.0041           \\
$\epsilon_2$      & 0.036        & 0.026       & 0.040  &0.037&	0.028	&0.037 & 0.036 & 0.026 & 0.038       \\
$n_s$        & 0.964        & 0.960       & 0.966    &0.963 &	0.961 &	0.967 & 0.963 & 0.961 & 0.967           \\
$r$         & 0.004        & 0.002       & 0.068   & 0.002&	0.002 &	0.055 & 0.002 & 0.002 & 0.065           \\
\hline
\end{tabular}
\caption{1 node secondary \textit{maxima a posteriori} values and 68\% confidence intervals of the feature and primordial parameters when a minimizer method is used.}
\end{table*}

\begin{table*}
\begin{tabular}{|r|ccc|ccc|}
\hline
Number of nodes $i$  & \multicolumn{6}{c|}{\textbf{2 training nodes} (other fits)}                                                    \\
\hline
\hline
\multicolumn{1}{|c|}{Modes}  & \multicolumn{3}{c|}{B.1 ($\Delta \chi ^2 = -16.1$)}  & \multicolumn{3}{c|}{B.2 ($\Delta \chi ^2 = -17.3$)}                           \\
\hline
 \multicolumn{1}{|c|}{Parameters} & Center  & lower & upper & Center & lower & upper \\
\hline
$\log_{10}l$    & -1.522 & -1.922 & 0.292   & -0.999 & -1.462 & -0.618 \\
$\log_{10}(|\tau_1|)$  & 2.565  & 2.296  & 3.297   & 2.896  & 2.400  & 3.200 \\
$\log_{10}(|u_1|)$     & -2.230 & -3.987 & -0.404 & -1.210 & -3.997 & -0.797 \\
$\log_{10}(|\tau_2|)$  & 2.277  & 2.213  & 2.300  & 2.002  & 1.970  & 2.033 5\\
$\log_{10}(|u_2|)$     & 1.452 & -3.895 & -1.424 & -1.693 & -2.240 & -1.426 \\
$\log_{10}(|u|_\mathrm{max})$      & -1.591 & -2.299 & -0.385 & -1.169 & -1.980 & -0.803\\
$\log_{10}(|s|_\mathrm{max})$      & -0.613 & -1.279 & -0.381 & -0.468 & -1.226 & -0.318\\
$\epsilon_1$       & 0.000  & 0.000  & 0.005   & 0.000  & 0.000  & 0.005 \\
$\epsilon_2$       & 0.037  & 0.026  & 0.037  & 0.036  & 0.026  & 0.039\\
$n_s$         & 0.963  & 0.961  & 0.967  & 0.964  & 0.960  & 0.968  \\
$r$          & 0.003  & 0.002  & 0.076 & 0.005  & 0.002  & 0.082 \\
\hline
\end{tabular}
\caption{2 nodes secondary \textit{maxima a posteriori} values and 68\% confidence intervals of the feature and primordial parameters, when a minimizer method is used.}
\end{table*}

\begin{table*}
\begin{tabular}{|c|ccc|ccc|ccc|ccc|ccc|}
\hline
Number of nodes $i$ & \multicolumn{15}{c|}{\textbf{3 training nodes} (other fits)} \\                                                             \hline                                     
\hline
Modes       & \multicolumn{3}{c|}{C.0 ($\Delta \chi ^2 = -21.4$)} & \multicolumn{3}{c|}{C.1 ($\Delta \chi ^2 = -20.7$)}    & \multicolumn{3}{c|}{C.3 ($\Delta \chi ^2 = -24.5$)}   & \multicolumn{3}{c|}{C.4 ($\Delta \chi ^2 = -22.3$)}   & \multicolumn{3}{c|}{C.5 ($\Delta \chi ^2 = -17.1$)}   \\ \hline
parameters  & Center      & lower & upper & Center      & lower & upper & Center      & lower & upper & Center      & lower & upper & Center      & lower & upper \\ \hline
$\log_{10}l$     & -0.979      & -1.634      & -0.559  & -1.166 & -1.514 & -0.646       & -1.208 & -1.209 & -0.821 & -1.297 & -1.475 & -0.721 & -0.909 & -1.443 & -0.690 \\ 
$\log_{10}(|\tau_1|)$ & 2.940       & 2.214       & 3,300  & 2.253  & 2.214  & 3.286   & 3.002  & 2.840  & 2.998  & 2.580  & 2.405  & 2.599  & 2.903  & 2.814  & 2.997  \\ 
$\log_{10}(|u_1|)$  & -3.205      & -4.000      & -0.825    & -1.658 & -3.989 & -0.939 & -1.974 & -1.952 & -1.288 & -2.169 & -3.931 & -1.507 & -1.205 & -3.953 & -0.971 \\ 
$\log_{10}(|\tau_2|)$ & 2.309       & 1.969       & 3.261 & 2.080  & 1.969  & 2.100   & 2.356  & 2.237  & 2.395  & 2.019  & 2.014  & 2.506  & 2.078  & 1.981  & 2.100  \\ 
$\log_{10}(|u_2|)$  & -1.486      & -3.998      & -1.120    & -3.487 & -3.965 & -1.572 & -2.216 & -3.724 & -1.965 & -2.741 & -3.303 & -1.560 & -2.735 & -3.965 & -1.572 \\ 
$\log_{10}(|\tau_3|)$ & 2.027       & 1.846       & 2.318  & 2.003  & 1.901  & 2.064 & 1.970  & 1.963  & 2.018  & 1.846  & 1.846  & 2.065  & 1.945  & 1.915  & 2.064  \\ 
$\log_{10}(|u_3|)$ & -1.822      & -3.261      & -1.415    & -2.185 & -2.860 & -1.559 & -1.733 & -2.019 & -1.565 & -1.807 & -3.261 & -1.513 & -2.296 & -2.822 & -1.629 \\ 
$\log_{10}(|u|_\mathrm{max})$  & -1.462      & -2.046      & -0.473     & -1.633 & -1.878 & -0.939 & -1.697 & -1.697 & -1.288 & -1.584 & -1.844 & -1.409 & -1.205 & -1.878 & -0.971 \\ 
$\log_{10}(|s|_\mathrm{max})$   & -0.768      & -1.287      & -0.206   & -0.925 & -1.216 & -0.231 & -0.808 & -0.994 & -0.516 & -0.872 & -1.151 & -0.410 & -0.547 & -1.216 & -0.231 \\ 
$\epsilon_1$     & 0.000       & 0.000       & 0.005    & 0.002  & 0.000  & 0.005  & 0.001  & 0.001  & 0.003  & 0.000  & 0.000  & 0.002  & 0.002  & 0.000  & 0.005  \\ 
$\epsilon_2$     & 0.036       & 0.024       & 0.039     & 0.033  & 0.025  & 0.039  & 0.035  & 0.031  & 0.036  & 0.036  & 0.031  & 0.039  & 0.033  & 0.025  & 0.038  \\ 
$n_s$     & 0.963       & 0.960       & 0.968       & 0.964  & 0.960  & 0.968  & 0.963  & 0.962  & 0.964  & 0.964  & 0.961  & 0.965  & 0.963  & 0.961  & 0.966  \\ 
$r$       & 0.004       & 0.002       & 0.087    & 0.025  & 0.002  & 0.082   & 0.018  & 0.018  & 0.044  & 0.003  & 0.002  & 0.039  & 0.028  & 0.002  & 0.081  \\ \hline
\end{tabular}
\caption{3 nodes secondary \textit{maxima a posteriori} values and 68\% confidence intervals of the feature and primordial parameters, when a minimizer method is used.}
\end{table*}

%FOUR MODES
\begin{table*}
\begin{tabular}{|r|ccc|ccc|}
\hline
\multicolumn{1}{|c|}{Number of nodes $i$}          & \multicolumn{6}{|c|}{\textbf{4 training nodes} (other fits)} \\
\hline         
\hline
\multicolumn{1}{|c|}{Modes}  &  \multicolumn{3}{c|}{D.1  ($\Delta \chi ^2 = -26.3$)} & \multicolumn{3}{c|}{D.2  ($\Delta \chi ^2 = -25.4$)}          \\
\hline
 \multicolumn{1}{|c|}{Parameters}  & Center           & lower      & upper  & Center           & lower      & upper  \\
\hline
$\log_{10}l$     & -1.294857    &   -1.211 & -0.751 &  -0.994857	& -1.522	& -0.613  \\
$\log_{10}(|\tau_1|)$  & 3.192       & 2.934 & 3.298 &  3.191	& 2.928 & 3.293       \\
$\log_{10}(|u_1|)$      & -1.978718905 &-1.952 &-0.851 &  -1.979 &	-1.952& -0.851     \\
$\log_{10}(|\tau_2|)$  & 2.99835       & 2.840 & 3.150 &  2.565	& 2.296	& 2.745       \\
$\log_{10}(|u_2|)$      & -1.95155 & -3.294 & -1.120 &  -2.230	& -3.987	& -0.404  \\
$\log_{10}(|\tau_3|)$  & 2.304      & 2.201  & 2.323 &  2.305 & 2.203 & 2.326       \\
$\log_{10}(|u_3|)$      & -1.528       & -2.654 & -0.262 &  -1.528	& -2.662 & -0.261  \\
$\log_{10}(|\tau_4|)$  & 2.002       & 1.951  & 2.297 &  2.002	& 1.951  & 2.297      \\
$\log_{10}(|u_4|)$      & -1.765      & -2.691 & -0.270 &  -1.763	 & -2.691 & -0.272  \\
$\log_{10}(|u|_\mathrm{max})$     & -1.528 &   -2.223 & -1.312 &  -1.511	& -2.334	& -1.422\\
$\log_{10}(|s|_\mathrm{max})$    &  -0.484 & -0.998 & -0.196 &  -0.845	& -1.322 & -0.231  \\
$\epsilon_1$     & 0.000        & 0.000       & 0.004  & 0.000	& 0.000	&0.005            \\
$\epsilon_2$      & 0.035        & 0.024       & 0.040  & 0.036 &	0.027	& 0.036       \\
$n_s$        & 0.963        & 0.961       & 0.966    & 0.964 &	0.960 &	0.967           \\
$r$         & 0.004        & 0.002       & 0.062   & 0.002&	0.002 &	0.060           \\
\hline
\end{tabular}
\caption{4 node secondary \textit{maxima a posteriori} values and 68\% confidence intervals of the feature and primordial parameters when a minimizer method is used.}
\end{table*}

%SUBSTRUCTURE
\begin{table*}
\begin{tabular}{|r|ccc|ccc|ccc|}
\hline
  & \multicolumn{9}{c|}{\textbf{MAP with substructure}}                                                    \\
\hline
\hline
\multicolumn{1}{|c|}{Modes}  & \multicolumn{3}{c|}{S.1 ($\Delta \chi ^2 = -19.6$)}  & \multicolumn{3}{c|}{S.2 ($\Delta \chi ^2 = -15.9$)}  & \multicolumn{3}{c|}{S.3 ($\Delta \chi ^2 = -24.1$)}                          \\
\hline
 \multicolumn{1}{|c|}{Parameters} & Center  & lower & upper & Center      & lower & upper & Center & lower & upper\\
\hline
$\log_{10}l$   &      1.228 & -1.415 & -0.855 & -1.101 & -1.999 & 0.499 & -1.098 & -1.939 & 0.501\\
$\log_{10}(|\tau_1|)$  & 2.062  & 1.993  & 2.100  &  3.180 & 2.998 & 3.400 &  3.180 & 2.999 & 3.400\\
$\log_{10}(|u_1|)$     & -1.900 & -3.907 & -1.555 &  -1.048 & -2.987 & -0.390&  -1.047 & -2.950 & -0.394\\
$\log_{10}(|\tau_2|)$  & 1.936  & 1.901  & 2.028  &  2.985 & 2.970 & 3.395&  2.986 & 2.967 & 3.391\\
$\log_{10}(|u_2|)$    & -1.937 & -3.178 & -1.605 &  -1.001 & -2.973 & -0.365&  -1.000 & -2.974 & -0.361\\
$\log_{10}(|\tau_3|)$           & -           & -          & -          & -           & -           & -     &    2.299  & 2.202  & 2.300       \\
$\log_{10}(|u_3|)$           & -           & -          & -          & -           & -           & -          &   -1.682 & -2.304 & -1.410     \\
$\log_{10}(|\tau_4|)$           & -           & -          & -          & -           & -           & -     &    2.003  & 1.961  & 2.030       \\
$\log_{10}(|u_4|)$           & -           & -          & -          & -           & -           & -          &   -1.749 & -2.039 & -1.495     \\
$\log_{10}(|u|_\mathrm{max})$  & -1.610 & -1.875 & -1.451& -0.876 & -2.758 & -0.355& -0.908 & -2.758 & -0.355\\
$\log_{10}(|s|_\mathrm{max})$ & -0.799 & -1.170 & -0.420 & -1.414 & -1.775 & -0.334& -0.138 & -1.775 & -0.334\\
$\epsilon_1$     & 0.001  & 0.000  & 0.005  & 0.0002	& 0.0001 & 0.0035 & 0.0001	& 0.0000 & 0.0036\\
$\epsilon_2$    & 0.034  & 0.028  & 0.037  & 0.036 & 0.029 & 0.038& 0.034 & 0.024 & 0.037\\
$n_s$ & 0.964       & 0.960      & 0.968 & 0.965  & 0.962  & 0.967   &  0.964	& 0.961	& 0.967\\
$r$ & 0.002       & 0.002      & 0.095 & 0.010  & 0.002  & 0.074  &   0.004 & 0.002	& 0.056\\
\hline
\end{tabular}
\caption{\textit{Maxima a posteriori} values, for those nodes which show some degree of sub-structure, and 68\% confidence intervals of the feature and primordial parameters, when a minimizer method is used.}
\end{table*}

%%%%%%%%%%%%%%%%%%%%%%%%%%%%%%%%%%%%%%%%%%%%%%%%%%%%%%%%%%

\bibliography{features}% Produces the bibliography via BibTeX.

\begin{thebibliography}{48}%
\makeatletter
\providecommand \@ifxundefined [1]{%
 \@ifx{#1\undefined}
}%
\providecommand \@ifnum [1]{%
 \ifnum #1\expandafter \@firstoftwo
 \else \expandafter \@secondoftwo
 \fi
}%
\providecommand \@ifx [1]{%
 \ifx #1\expandafter \@firstoftwo
 \else \expandafter \@secondoftwo
 \fi
}%
\providecommand \natexlab [1]{#1}%
\providecommand \enquote  [1]{``#1''}%
\providecommand \bibnamefont  [1]{#1}%
\providecommand \bibfnamefont [1]{#1}%
\providecommand \citenamefont [1]{#1}%
\providecommand \href@noop [0]{\@secondoftwo}%
\providecommand \href [0]{\begingroup \@sanitize@url \@href}%
\providecommand \@href[1]{\@@startlink{#1}\@@href}%
\providecommand \@@href[1]{\endgroup#1\@@endlink}%
\providecommand \@sanitize@url [0]{\catcode `\\12\catcode `\$12\catcode
  `\&12\catcode `\#12\catcode `\^12\catcode `\_12\catcode `\%12\relax}%
\providecommand \@@startlink[1]{}%
\providecommand \@@endlink[0]{}%
\providecommand \url  [0]{\begingroup\@sanitize@url \@url }%
\providecommand \@url [1]{\endgroup\@href {#1}{\urlprefix }}%
\providecommand \urlprefix  [0]{URL }%
\providecommand \Eprint [0]{\href }%
\providecommand \doibase [0]{http://dx.doi.org/}%
\providecommand \selectlanguage [0]{\@gobble}%
\providecommand \bibinfo  [0]{\@secondoftwo}%
\providecommand \bibfield  [0]{\@secondoftwo}%
\providecommand \translation [1]{[#1]}%
\providecommand \BibitemOpen [0]{}%
\providecommand \bibitemStop [0]{}%
\providecommand \bibitemNoStop [0]{.\EOS\space}%
\providecommand \EOS [0]{\spacefactor3000\relax}%
\providecommand \BibitemShut  [1]{\csname bibitem#1\endcsname}%
\let\auto@bib@innerbib\@empty
%</preamble>
\bibitem [{\citenamefont {{Chluba}}\ \emph {et~al.}(2015)\citenamefont
  {{Chluba}}, \citenamefont {{Hamann}},\ and\ \citenamefont
  {{Patil}}}]{FeaturesReview}%
  \BibitemOpen
  \bibfield  {author} {\bibinfo {author} {\bibfnamefont {J.}~\bibnamefont
  {{Chluba}}}, \bibinfo {author} {\bibfnamefont {J.}~\bibnamefont {{Hamann}}},
  \ and\ \bibinfo {author} {\bibfnamefont {S.~P.}\ \bibnamefont {{Patil}}},\
  }\href {\doibase 10.1142/S0218271815300232} {\bibfield  {journal} {\bibinfo
  {journal} {International Journal of Modern Physics D}\ }\textbf {\bibinfo
  {volume} {24}},\ \bibinfo {eid} {1530023} (\bibinfo {year} {2015})},\ \Eprint
  {http://arxiv.org/abs/1505.01834} {arXiv:1505.01834} \BibitemShut {NoStop}%
\bibitem [{\citenamefont {{Chen}}(2010)}]{2010AdAst2010E..72C}%
  \BibitemOpen
  \bibfield  {author} {\bibinfo {author} {\bibfnamefont {X.}~\bibnamefont
  {{Chen}}},\ }\href {\doibase 10.1155/2010/638979} {\bibfield  {journal}
  {\bibinfo  {journal} {Advances in Astronomy}\ }\textbf {\bibinfo {volume}
  {2010}},\ \bibinfo {eid} {638979} (\bibinfo {year} {2010})},\ \Eprint
  {http://arxiv.org/abs/1002.1416} {arXiv:1002.1416 [astro-ph.CO]} \BibitemShut
  {NoStop}%
\bibitem [{\citenamefont {Slosar}\ \emph {et~al.}(2019)\citenamefont {Slosar},
  \citenamefont {Chen}, \citenamefont {Dvorkin}, \citenamefont {Meerburg},
  \citenamefont {Wallisch}, \citenamefont {Green},\ and\ \citenamefont
  {Silverstein}}]{Slosar2019Scratches}%
  \BibitemOpen
  \bibfield  {author} {\bibinfo {author} {\bibfnamefont {A.}~\bibnamefont
  {Slosar}}, \bibinfo {author} {\bibfnamefont {X.}~\bibnamefont {Chen}},
  \bibinfo {author} {\bibfnamefont {C.}~\bibnamefont {Dvorkin}}, \bibinfo
  {author} {\bibfnamefont {D.}~\bibnamefont {Meerburg}}, \bibinfo {author}
  {\bibfnamefont {B.}~\bibnamefont {Wallisch}}, \bibinfo {author}
  {\bibfnamefont {D.}~\bibnamefont {Green}}, \ and\ \bibinfo {author}
  {\bibfnamefont {E.}~\bibnamefont {Silverstein}},\ }\href
  {https://baas.aas.org/pub/2020n3i098} {\bibfield  {journal} {\bibinfo
  {journal} {Bulletin of the AAS}\ }\textbf {\bibinfo {volume} {51}} (\bibinfo
  {year} {2019})},\ \bibinfo {note}
  {https://baas.aas.org/pub/2020n3i098}\BibitemShut {NoStop}%
\bibitem [{\citenamefont {Palma}(2015)}]{Palma:2014hra}%
  \BibitemOpen
  \bibfield  {author} {\bibinfo {author} {\bibfnamefont {G.~A.}\ \bibnamefont
  {Palma}},\ }\href {\doibase 10.1088/1475-7516/2015/04/035} {\bibfield
  {journal} {\bibinfo  {journal} {JCAP}\ }\textbf {\bibinfo {volume} {04}},\
  \bibinfo {pages} {035} (\bibinfo {year} {2015})},\ \Eprint
  {http://arxiv.org/abs/1412.5615} {arXiv:1412.5615 [hep-th]} \BibitemShut
  {NoStop}%
\bibitem [{\citenamefont {{Ach{\'u}carro}}\ \emph {et~al.}(2013)\citenamefont
  {{Ach{\'u}carro}}, \citenamefont {{Gong}}, \citenamefont {{Palma}},\ and\
  \citenamefont {{Patil}}}]{2013PhRvD..87l1301A}%
  \BibitemOpen
  \bibfield  {author} {\bibinfo {author} {\bibfnamefont {A.}~\bibnamefont
  {{Ach{\'u}carro}}}, \bibinfo {author} {\bibfnamefont {J.-O.}\ \bibnamefont
  {{Gong}}}, \bibinfo {author} {\bibfnamefont {G.~A.}\ \bibnamefont {{Palma}}},
  \ and\ \bibinfo {author} {\bibfnamefont {S.~P.}\ \bibnamefont {{Patil}}},\
  }\href {\doibase 10.1103/PhysRevD.87.121301} {\bibfield  {journal} {\bibinfo
  {journal} {\prd}\ }\textbf {\bibinfo {volume} {87}},\ \bibinfo {eid} {121301}
  (\bibinfo {year} {2013})},\ \Eprint {http://arxiv.org/abs/1211.5619}
  {arXiv:1211.5619 [astro-ph.CO]} \BibitemShut {NoStop}%
\bibitem [{\citenamefont {{Planck Collaboration}}\ \emph
  {et~al.}(2018{\natexlab{a}})\citenamefont {{Planck Collaboration}},
  \citenamefont {{Aghanim}}, \citenamefont {{Akrami}}, \citenamefont
  {{Ashdown}}, \citenamefont {{Aumont}}, \citenamefont {{Baccigalupi}},
  \citenamefont {{Ballardini}}, \citenamefont {{Banday}}, \citenamefont
  {{Barreiro}},\ and\ \citenamefont {{Bartolo}}}]{2018arXiv180706209P}%
  \BibitemOpen
  \bibfield  {author} {\bibinfo {author} {\bibnamefont {{Planck
  Collaboration}}}, \bibinfo {author} {\bibfnamefont {N.}~\bibnamefont
  {{Aghanim}}}, \bibinfo {author} {\bibfnamefont {Y.}~\bibnamefont {{Akrami}}},
  \bibinfo {author} {\bibfnamefont {M.}~\bibnamefont {{Ashdown}}}, \bibinfo
  {author} {\bibfnamefont {J.}~\bibnamefont {{Aumont}}}, \bibinfo {author}
  {\bibfnamefont {C.}~\bibnamefont {{Baccigalupi}}}, \bibinfo {author}
  {\bibfnamefont {M.}~\bibnamefont {{Ballardini}}}, \bibinfo {author}
  {\bibfnamefont {A.~J.}\ \bibnamefont {{Banday}}}, \bibinfo {author}
  {\bibfnamefont {R.~B.}\ \bibnamefont {{Barreiro}}}, \ and\ \bibinfo {author}
  {\bibfnamefont {N.}~\bibnamefont {{Bartolo}}},\ }\href@noop {} {\bibfield
  {journal} {\bibinfo  {journal} {arXiv e-prints}\ ,\ \bibinfo {eid}
  {arXiv:1807.06209}} (\bibinfo {year} {2018}{\natexlab{a}})},\ \Eprint
  {http://arxiv.org/abs/1807.06209} {arXiv:1807.06209 [astro-ph.CO]}
  \BibitemShut {NoStop}%
\bibitem [{\citenamefont {{Planck Collaboration}}\ \emph
  {et~al.}(2018{\natexlab{b}})\citenamefont {{Planck Collaboration}},
  \citenamefont {{Akrami}}, \citenamefont {{Arroja}}, \citenamefont
  {{Ashdown}}, \citenamefont {{Aumont}}, \citenamefont {{Baccigalupi}},
  \citenamefont {{Ballardini}}, \citenamefont {{Banday}}, \citenamefont
  {{Barreiro}},\ and\ \citenamefont {{Bartolo}}}]{2018arXiv180706211P}%
  \BibitemOpen
  \bibfield  {author} {\bibinfo {author} {\bibnamefont {{Planck
  Collaboration}}}, \bibinfo {author} {\bibfnamefont {Y.}~\bibnamefont
  {{Akrami}}}, \bibinfo {author} {\bibfnamefont {F.}~\bibnamefont {{Arroja}}},
  \bibinfo {author} {\bibfnamefont {M.}~\bibnamefont {{Ashdown}}}, \bibinfo
  {author} {\bibfnamefont {J.}~\bibnamefont {{Aumont}}}, \bibinfo {author}
  {\bibfnamefont {C.}~\bibnamefont {{Baccigalupi}}}, \bibinfo {author}
  {\bibfnamefont {M.}~\bibnamefont {{Ballardini}}}, \bibinfo {author}
  {\bibfnamefont {A.~J.}\ \bibnamefont {{Banday}}}, \bibinfo {author}
  {\bibfnamefont {R.~B.}\ \bibnamefont {{Barreiro}}}, \ and\ \bibinfo {author}
  {\bibfnamefont {N.}~\bibnamefont {{Bartolo}}},\ }\href@noop {} {\bibfield
  {journal} {\bibinfo  {journal} {arXiv e-prints}\ ,\ \bibinfo {eid}
  {arXiv:1807.06211}} (\bibinfo {year} {2018}{\natexlab{b}})},\ \Eprint
  {http://arxiv.org/abs/1807.06211} {arXiv:1807.06211 [astro-ph.CO]}
  \BibitemShut {NoStop}%
\bibitem [{\citenamefont {{Ach{\'u}carro}}\ \emph
  {et~al.}(2014{\natexlab{a}})\citenamefont {{Ach{\'u}carro}}, \citenamefont
  {{Atal}}, \citenamefont {{Ortiz}},\ and\ \citenamefont
  {{Torrado}}}]{2014PhRvD..89j3006A}%
  \BibitemOpen
  \bibfield  {author} {\bibinfo {author} {\bibfnamefont {A.}~\bibnamefont
  {{Ach{\'u}carro}}}, \bibinfo {author} {\bibfnamefont {V.}~\bibnamefont
  {{Atal}}}, \bibinfo {author} {\bibfnamefont {P.}~\bibnamefont {{Ortiz}}}, \
  and\ \bibinfo {author} {\bibfnamefont {J.}~\bibnamefont {{Torrado}}},\ }\href
  {\doibase 10.1103/PhysRevD.89.103006} {\bibfield  {journal} {\bibinfo
  {journal} {\prd}\ }\textbf {\bibinfo {volume} {89}},\ \bibinfo {eid} {103006}
  (\bibinfo {year} {2014}{\natexlab{a}})},\ \Eprint
  {http://arxiv.org/abs/1311.2552} {arXiv:1311.2552} \BibitemShut {NoStop}%
\bibitem [{\citenamefont {{Ach{\'u}carro}}\ \emph
  {et~al.}(2014{\natexlab{b}})\citenamefont {{Ach{\'u}carro}}, \citenamefont
  {{Atal}}, \citenamefont {{Hu}}, \citenamefont {{Ortiz}},\ and\ \citenamefont
  {{Torrado}}}]{2014PhRvD..90b3511A}%
  \BibitemOpen
  \bibfield  {author} {\bibinfo {author} {\bibfnamefont {A.}~\bibnamefont
  {{Ach{\'u}carro}}}, \bibinfo {author} {\bibfnamefont {V.}~\bibnamefont
  {{Atal}}}, \bibinfo {author} {\bibfnamefont {B.}~\bibnamefont {{Hu}}},
  \bibinfo {author} {\bibfnamefont {P.}~\bibnamefont {{Ortiz}}}, \ and\
  \bibinfo {author} {\bibfnamefont {J.}~\bibnamefont {{Torrado}}},\ }\href
  {\doibase 10.1103/PhysRevD.90.023511} {\bibfield  {journal} {\bibinfo
  {journal} {\prd}\ }\textbf {\bibinfo {volume} {90}},\ \bibinfo {eid} {023511}
  (\bibinfo {year} {2014}{\natexlab{b}})},\ \Eprint
  {http://arxiv.org/abs/1404.7522} {arXiv:1404.7522} \BibitemShut {NoStop}%
\bibitem [{\citenamefont {Hu}\ and\ \citenamefont
  {Torrado}(2015)}]{Hu:2014hra}%
  \BibitemOpen
  \bibfield  {author} {\bibinfo {author} {\bibfnamefont {B.}~\bibnamefont
  {Hu}}\ and\ \bibinfo {author} {\bibfnamefont {J.}~\bibnamefont {Torrado}},\
  }\href {\doibase 10.1103/PhysRevD.91.064039} {\bibfield  {journal} {\bibinfo
  {journal} {Phys.\ Rev.\ D}\ }\textbf {\bibinfo {volume} {91}},\ \bibinfo
  {pages} {064039} (\bibinfo {year} {2015})},\ \Eprint
  {http://arxiv.org/abs/1410.4804} {arXiv:1410.4804 [astro-ph.CO]} \BibitemShut
  {NoStop}%
\bibitem [{\citenamefont {{Torrado}}\ \emph {et~al.}(2017)\citenamefont
  {{Torrado}}, \citenamefont {{Hu}},\ and\ \citenamefont
  {{Ach{\'u}carro}}}]{2017PhRvD..96h3515T}%
  \BibitemOpen
  \bibfield  {author} {\bibinfo {author} {\bibfnamefont {J.}~\bibnamefont
  {{Torrado}}}, \bibinfo {author} {\bibfnamefont {B.}~\bibnamefont {{Hu}}}, \
  and\ \bibinfo {author} {\bibfnamefont {A.}~\bibnamefont {{Ach{\'u}carro}}},\
  }\href {\doibase 10.1103/PhysRevD.96.083515} {\bibfield  {journal} {\bibinfo
  {journal} {\prd}\ }\textbf {\bibinfo {volume} {96}},\ \bibinfo {eid} {083515}
  (\bibinfo {year} {2017})},\ \Eprint {http://arxiv.org/abs/1611.10350}
  {arXiv:1611.10350} \BibitemShut {NoStop}%
\bibitem [{\citenamefont {{Planck Collaboration}}\ \emph
  {et~al.}(2016)\citenamefont {{Planck Collaboration}}, \citenamefont {{Ade}},
  \citenamefont {{Aghanim}}, \citenamefont {{Arnaud}}, \citenamefont
  {{Arroja}}, \citenamefont {{Ashdown}}, \citenamefont {{Aumont}},
  \citenamefont {{Baccigalupi}}, \citenamefont {{Ballardini}}, \citenamefont
  {{Banday}},\ and\ \citenamefont {et~al.}}]{2016A&A...594A..20P}%
  \BibitemOpen
  \bibfield  {author} {\bibinfo {author} {\bibnamefont {{Planck
  Collaboration}}}, \bibinfo {author} {\bibfnamefont {P.~A.~R.}\ \bibnamefont
  {{Ade}}}, \bibinfo {author} {\bibfnamefont {N.}~\bibnamefont {{Aghanim}}},
  \bibinfo {author} {\bibfnamefont {M.}~\bibnamefont {{Arnaud}}}, \bibinfo
  {author} {\bibfnamefont {F.}~\bibnamefont {{Arroja}}}, \bibinfo {author}
  {\bibfnamefont {M.}~\bibnamefont {{Ashdown}}}, \bibinfo {author}
  {\bibfnamefont {J.}~\bibnamefont {{Aumont}}}, \bibinfo {author}
  {\bibfnamefont {C.}~\bibnamefont {{Baccigalupi}}}, \bibinfo {author}
  {\bibfnamefont {M.}~\bibnamefont {{Ballardini}}}, \bibinfo {author}
  {\bibfnamefont {A.~J.}\ \bibnamefont {{Banday}}}, \ and\ \bibinfo {author}
  {\bibnamefont {et~al.}},\ }\href {\doibase 10.1051/0004-6361/201525898}
  {\bibfield  {journal} {\bibinfo  {journal} {Astronomy and Astrophysics}\
  }\textbf {\bibinfo {volume} {594}},\ \bibinfo {eid} {A20} (\bibinfo {year}
  {2016})},\ \Eprint {http://arxiv.org/abs/1502.02114} {arXiv:1502.02114}
  \BibitemShut {NoStop}%
\bibitem [{\citenamefont {{Hazra}}\ \emph {et~al.}(2013)\citenamefont
  {{Hazra}}, \citenamefont {{Shafieloo}},\ and\ \citenamefont
  {{Souradeep}}}]{2013JCAP...07..031H}%
  \BibitemOpen
  \bibfield  {author} {\bibinfo {author} {\bibfnamefont {D.~K.}\ \bibnamefont
  {{Hazra}}}, \bibinfo {author} {\bibfnamefont {A.}~\bibnamefont
  {{Shafieloo}}}, \ and\ \bibinfo {author} {\bibfnamefont {T.}~\bibnamefont
  {{Souradeep}}},\ }\href {\doibase 10.1088/1475-7516/2013/07/031} {\bibfield
  {journal} {\bibinfo  {journal} {Journal of Cosmology and Astroparticle
  Physics}\ }\textbf {\bibinfo {volume} {2013}},\ \bibinfo {eid} {031}
  (\bibinfo {year} {2013})},\ \Eprint {http://arxiv.org/abs/1303.4143}
  {arXiv:1303.4143 [astro-ph.CO]} \BibitemShut {NoStop}%
\bibitem [{\citenamefont {{Hunt}}\ and\ \citenamefont
  {{Sarkar}}(2014)}]{2014JCAP...01..025H}%
  \BibitemOpen
  \bibfield  {author} {\bibinfo {author} {\bibfnamefont {P.}~\bibnamefont
  {{Hunt}}}\ and\ \bibinfo {author} {\bibfnamefont {S.}~\bibnamefont
  {{Sarkar}}},\ }\href {\doibase 10.1088/1475-7516/2014/01/025} {\bibfield
  {journal} {\bibinfo  {journal} {Journal of Cosmology and Astroparticle
  Physics}\ }\textbf {\bibinfo {volume} {2014}},\ \bibinfo {eid} {025}
  (\bibinfo {year} {2014})},\ \Eprint {http://arxiv.org/abs/1308.2317}
  {arXiv:1308.2317 [astro-ph.CO]} \BibitemShut {NoStop}%
\bibitem [{\citenamefont {{Ravenni}}\ \emph {et~al.}(2016)\citenamefont
  {{Ravenni}}, \citenamefont {{Verde}},\ and\ \citenamefont
  {{Cuesta}}}]{2016JCAP...08..028R}%
  \BibitemOpen
  \bibfield  {author} {\bibinfo {author} {\bibfnamefont {A.}~\bibnamefont
  {{Ravenni}}}, \bibinfo {author} {\bibfnamefont {L.}~\bibnamefont {{Verde}}},
  \ and\ \bibinfo {author} {\bibfnamefont {A.~J.}\ \bibnamefont {{Cuesta}}},\
  }\href {\doibase 10.1088/1475-7516/2016/08/028} {\bibfield  {journal}
  {\bibinfo  {journal} {Journal of Cosmology and Astroparticle Physics}\
  }\textbf {\bibinfo {volume} {2016}},\ \bibinfo {eid} {028} (\bibinfo {year}
  {2016})},\ \Eprint {http://arxiv.org/abs/1605.06637} {arXiv:1605.06637
  [astro-ph.CO]} \BibitemShut {NoStop}%
\bibitem [{\citenamefont {{Durakovic}}\ \emph {et~al.}(2018)\citenamefont
  {{Durakovic}}, \citenamefont {{Hunt}}, \citenamefont {{Mukherjee}},
  \citenamefont {{Sarkar}},\ and\ \citenamefont
  {{Souradeep}}}]{2018JCAP...02..012D}%
  \BibitemOpen
  \bibfield  {author} {\bibinfo {author} {\bibfnamefont {A.}~\bibnamefont
  {{Durakovic}}}, \bibinfo {author} {\bibfnamefont {P.}~\bibnamefont {{Hunt}}},
  \bibinfo {author} {\bibfnamefont {S.}~\bibnamefont {{Mukherjee}}}, \bibinfo
  {author} {\bibfnamefont {S.}~\bibnamefont {{Sarkar}}}, \ and\ \bibinfo
  {author} {\bibfnamefont {T.}~\bibnamefont {{Souradeep}}},\ }\href {\doibase
  10.1088/1475-7516/2018/02/012} {\bibfield  {journal} {\bibinfo  {journal}
  {Journal of Cosmology and Astroparticle Physics}\ }\textbf {\bibinfo {volume}
  {2018}},\ \bibinfo {eid} {012} (\bibinfo {year} {2018})},\ \Eprint
  {http://arxiv.org/abs/1711.08441} {arXiv:1711.08441 [astro-ph.CO]}
  \BibitemShut {NoStop}%
\bibitem [{\citenamefont {{Ballardini}}(2019)}]{2019PDU....23..245B}%
  \BibitemOpen
  \bibfield  {author} {\bibinfo {author} {\bibfnamefont {M.}~\bibnamefont
  {{Ballardini}}},\ }\href {\doibase 10.1016/j.dark.2018.11.006} {\bibfield
  {journal} {\bibinfo  {journal} {Physics of the Dark Universe}\ }\textbf
  {\bibinfo {volume} {23}},\ \bibinfo {eid} {100245} (\bibinfo {year}
  {2019})},\ \Eprint {http://arxiv.org/abs/1807.05521} {arXiv:1807.05521
  [astro-ph.CO]} \BibitemShut {NoStop}%
\bibitem [{\citenamefont {Durakovic}\ \emph {et~al.}(2019)\citenamefont
  {Durakovic}, \citenamefont {Hunt}, \citenamefont {Patil},\ and\ \citenamefont
  {Sarkar}}]{epsilon}%
  \BibitemOpen
  \bibfield  {author} {\bibinfo {author} {\bibfnamefont {A.}~\bibnamefont
  {Durakovic}}, \bibinfo {author} {\bibfnamefont {P.}~\bibnamefont {Hunt}},
  \bibinfo {author} {\bibfnamefont {S.~P.}\ \bibnamefont {Patil}}, \ and\
  \bibinfo {author} {\bibfnamefont {S.}~\bibnamefont {Sarkar}},\ }\href
  {\doibase 10.21468/SciPostPhys.7.4.049} {\bibfield  {journal} {\bibinfo
  {journal} {SciPost Phys.}\ }\textbf {\bibinfo {volume} {7}},\ \bibinfo
  {pages} {049} (\bibinfo {year} {2019})},\ \Eprint
  {http://arxiv.org/abs/1904.00991} {arXiv:1904.00991 [astro-ph.CO]}
  \BibitemShut {NoStop}%
%%CITATION = ARXIV:1904.00991;%%
\bibitem [{\citenamefont {{Handley}}\ \emph {et~al.}(2019)\citenamefont
  {{Handley}}, \citenamefont {{Lasenby}}, \citenamefont {{Peiris}},\ and\
  \citenamefont {{Hobson}}}]{2019PhRvD.100j3511H}%
  \BibitemOpen
  \bibfield  {author} {\bibinfo {author} {\bibfnamefont {W.~J.}\ \bibnamefont
  {{Handley}}}, \bibinfo {author} {\bibfnamefont {A.~N.}\ \bibnamefont
  {{Lasenby}}}, \bibinfo {author} {\bibfnamefont {H.~V.}\ \bibnamefont
  {{Peiris}}}, \ and\ \bibinfo {author} {\bibfnamefont {M.~P.}\ \bibnamefont
  {{Hobson}}},\ }\href {\doibase 10.1103/PhysRevD.100.103511} {\bibfield
  {journal} {\bibinfo  {journal} {\prd}\ }\textbf {\bibinfo {volume} {100}},\
  \bibinfo {eid} {103511} (\bibinfo {year} {2019})},\ \Eprint
  {http://arxiv.org/abs/1908.00906} {arXiv:1908.00906 [astro-ph.CO]}
  \BibitemShut {NoStop}%
\bibitem [{\citenamefont {{Appleby}}\ \emph {et~al.}(2016)\citenamefont
  {{Appleby}}, \citenamefont {{Gong}}, \citenamefont {{Hazra}}, \citenamefont
  {{Shafieloo}},\ and\ \citenamefont {{Sypsas}}}]{2016PhLB..760..297A}%
  \BibitemOpen
  \bibfield  {author} {\bibinfo {author} {\bibfnamefont {S.}~\bibnamefont
  {{Appleby}}}, \bibinfo {author} {\bibfnamefont {J.-O.}\ \bibnamefont
  {{Gong}}}, \bibinfo {author} {\bibfnamefont {D.~K.}\ \bibnamefont {{Hazra}}},
  \bibinfo {author} {\bibfnamefont {A.}~\bibnamefont {{Shafieloo}}}, \ and\
  \bibinfo {author} {\bibfnamefont {S.}~\bibnamefont {{Sypsas}}},\ }\href
  {\doibase 10.1016/j.physletb.2016.07.004} {\bibfield  {journal} {\bibinfo
  {journal} {Physics Letters B}\ }\textbf {\bibinfo {volume} {760}},\ \bibinfo
  {pages} {297} (\bibinfo {year} {2016})},\ \Eprint
  {http://arxiv.org/abs/1512.08977} {arXiv:1512.08977 [astro-ph.CO]}
  \BibitemShut {NoStop}%
\bibitem [{\citenamefont {{Handley}}\ and\ \citenamefont
  {{Millea}}(2019)}]{2019Entrp..21..272H}%
  \BibitemOpen
  \bibfield  {author} {\bibinfo {author} {\bibfnamefont {W.}~\bibnamefont
  {{Handley}}}\ and\ \bibinfo {author} {\bibfnamefont {M.}~\bibnamefont
  {{Millea}}},\ }\href {\doibase 10.3390/e21030272} {\bibfield  {journal}
  {\bibinfo  {journal} {Entropy}\ }\textbf {\bibinfo {volume} {21}},\ \bibinfo
  {pages} {272} (\bibinfo {year} {2019})},\ \Eprint
  {http://arxiv.org/abs/1804.08143} {arXiv:1804.08143 [math.ST]} \BibitemShut
  {NoStop}%
\bibitem [{\citenamefont {Gariazzo}\ and\ \citenamefont
  {Mena}(2019)}]{Gariazzo:2018meg}%
  \BibitemOpen
  \bibfield  {author} {\bibinfo {author} {\bibfnamefont {S.}~\bibnamefont
  {Gariazzo}}\ and\ \bibinfo {author} {\bibfnamefont {O.}~\bibnamefont
  {Mena}},\ }\href {\doibase 10.1103/PhysRevD.99.021301} {\bibfield  {journal}
  {\bibinfo  {journal} {Phys. Rev. D}\ }\textbf {\bibinfo {volume} {99}},\
  \bibinfo {pages} {021301} (\bibinfo {year} {2019})},\ \Eprint
  {http://arxiv.org/abs/1812.05449} {arXiv:1812.05449 [astro-ph.CO]}
  \BibitemShut {NoStop}%
\bibitem [{\citenamefont {Aghanim}\ \emph {et~al.}(2018)\citenamefont {Aghanim}
  \emph {et~al.}}]{Aghanim:2018eyx}%
  \BibitemOpen
  \bibfield  {author} {\bibinfo {author} {\bibfnamefont {N.}~\bibnamefont
  {Aghanim}} \emph {et~al.} (\bibinfo {collaboration} {Planck}),\ }\href@noop
  {} {\  (\bibinfo {year} {2018})},\ \Eprint {http://arxiv.org/abs/1807.06209}
  {arXiv:1807.06209 [astro-ph.CO]} \BibitemShut {NoStop}%
\bibitem [{\citenamefont {{Cheung}}\ \emph {et~al.}(2008)\citenamefont
  {{Cheung}}, \citenamefont {{Fitzpatrick}}, \citenamefont {{Kaplan}},
  \citenamefont {{Senatore}},\ and\ \citenamefont
  {{Creminelli}}}]{2008JHEP...03..014C}%
  \BibitemOpen
  \bibfield  {author} {\bibinfo {author} {\bibfnamefont {C.}~\bibnamefont
  {{Cheung}}}, \bibinfo {author} {\bibfnamefont {A.~L.}\ \bibnamefont
  {{Fitzpatrick}}}, \bibinfo {author} {\bibfnamefont {J.}~\bibnamefont
  {{Kaplan}}}, \bibinfo {author} {\bibfnamefont {L.}~\bibnamefont
  {{Senatore}}}, \ and\ \bibinfo {author} {\bibfnamefont {P.}~\bibnamefont
  {{Creminelli}}},\ }\href {\doibase 10.1088/1126-6708/2008/03/014} {\bibfield
  {journal} {\bibinfo  {journal} {Journal of High Energy Physics}\ }\textbf
  {\bibinfo {volume} {3}},\ \bibinfo {eid} {014-014} (\bibinfo {year}
  {2008})},\ \Eprint {http://arxiv.org/abs/0709.0293} {arXiv:0709.0293
  [hep-th]} \BibitemShut {NoStop}%
\bibitem [{\citenamefont {Achucarro}\ \emph {et~al.}(2012)\citenamefont
  {Achucarro}, \citenamefont {Gong}, \citenamefont {Hardeman}, \citenamefont
  {Palma},\ and\ \citenamefont {Patil}}]{Achucarro:2012sm}%
  \BibitemOpen
  \bibfield  {author} {\bibinfo {author} {\bibfnamefont {A.}~\bibnamefont
  {Achucarro}}, \bibinfo {author} {\bibfnamefont {J.-O.}\ \bibnamefont {Gong}},
  \bibinfo {author} {\bibfnamefont {S.}~\bibnamefont {Hardeman}}, \bibinfo
  {author} {\bibfnamefont {G.~A.}\ \bibnamefont {Palma}}, \ and\ \bibinfo
  {author} {\bibfnamefont {S.~P.}\ \bibnamefont {Patil}},\ }\href {\doibase
  10.1007/JHEP05(2012)066} {\bibfield  {journal} {\bibinfo  {journal} {JHEP}\
  }\textbf {\bibinfo {volume} {05}},\ \bibinfo {pages} {066} (\bibinfo {year}
  {2012})},\ \Eprint {http://arxiv.org/abs/1201.6342} {arXiv:1201.6342
  [hep-th]} \BibitemShut {NoStop}%
\bibitem [{\citenamefont {{Ach{\'u}carro}}\ \emph
  {et~al.}(2011{\natexlab{a}})\citenamefont {{Ach{\'u}carro}}, \citenamefont
  {{Gong}}, \citenamefont {{Hardeman}}, \citenamefont {{Palma}},\ and\
  \citenamefont {{Patil}}}]{2011JCAP...01..030A}%
  \BibitemOpen
  \bibfield  {author} {\bibinfo {author} {\bibfnamefont {A.}~\bibnamefont
  {{Ach{\'u}carro}}}, \bibinfo {author} {\bibfnamefont {J.-O.}\ \bibnamefont
  {{Gong}}}, \bibinfo {author} {\bibfnamefont {S.}~\bibnamefont {{Hardeman}}},
  \bibinfo {author} {\bibfnamefont {G.~A.}\ \bibnamefont {{Palma}}}, \ and\
  \bibinfo {author} {\bibfnamefont {S.~P.}\ \bibnamefont {{Patil}}},\ }\href
  {\doibase 10.1088/1475-7516/2011/01/030} {\bibfield  {journal} {\bibinfo
  {journal} {Journal of Cosmology and Astroparticle Physics}\ }\textbf
  {\bibinfo {volume} {1}},\ \bibinfo {eid} {030} (\bibinfo {year}
  {2011}{\natexlab{a}})},\ \Eprint {http://arxiv.org/abs/1010.3693}
  {arXiv:1010.3693 [hep-ph]} \BibitemShut {NoStop}%
\bibitem [{\citenamefont {{Ach{\'u}carro}}\ \emph
  {et~al.}(2011{\natexlab{b}})\citenamefont {{Ach{\'u}carro}}, \citenamefont
  {{Gong}}, \citenamefont {{Hardeman}}, \citenamefont {{Palma}},\ and\
  \citenamefont {{Patil}}}]{2011PhRvD..84d3502A}%
  \BibitemOpen
  \bibfield  {author} {\bibinfo {author} {\bibfnamefont {A.}~\bibnamefont
  {{Ach{\'u}carro}}}, \bibinfo {author} {\bibfnamefont {J.-O.}\ \bibnamefont
  {{Gong}}}, \bibinfo {author} {\bibfnamefont {S.}~\bibnamefont {{Hardeman}}},
  \bibinfo {author} {\bibfnamefont {G.~A.}\ \bibnamefont {{Palma}}}, \ and\
  \bibinfo {author} {\bibfnamefont {S.~P.}\ \bibnamefont {{Patil}}},\ }\href
  {\doibase 10.1103/PhysRevD.84.043502} {\bibfield  {journal} {\bibinfo
  {journal} {\prd}\ }\textbf {\bibinfo {volume} {84}},\ \bibinfo {eid} {043502}
  (\bibinfo {year} {2011}{\natexlab{b}})},\ \Eprint
  {http://arxiv.org/abs/1005.3848} {arXiv:1005.3848 [hep-th]} \BibitemShut
  {NoStop}%
\bibitem [{\citenamefont {{Ach{\'u}carro}}\ \emph {et~al.}(2012)\citenamefont
  {{Ach{\'u}carro}}, \citenamefont {{Atal}}, \citenamefont {{C{\'e}spedes}},
  \citenamefont {{Gong}}, \citenamefont {{Palma}},\ and\ \citenamefont
  {{Patil}}}]{EFTCS1}%
  \BibitemOpen
  \bibfield  {author} {\bibinfo {author} {\bibfnamefont {A.}~\bibnamefont
  {{Ach{\'u}carro}}}, \bibinfo {author} {\bibfnamefont {V.}~\bibnamefont
  {{Atal}}}, \bibinfo {author} {\bibfnamefont {S.}~\bibnamefont
  {{C{\'e}spedes}}}, \bibinfo {author} {\bibfnamefont {J.-O.}\ \bibnamefont
  {{Gong}}}, \bibinfo {author} {\bibfnamefont {G.~A.}\ \bibnamefont {{Palma}}},
  \ and\ \bibinfo {author} {\bibfnamefont {S.~P.}\ \bibnamefont {{Patil}}},\
  }\href {\doibase 10.1103/PhysRevD.86.121301} {\bibfield  {journal} {\bibinfo
  {journal} {\prd}\ }\textbf {\bibinfo {volume} {86}},\ \bibinfo {eid} {121301}
  (\bibinfo {year} {2012})}\BibitemShut {NoStop}%
\bibitem [{\citenamefont {{Weinberg}}(2005)}]{2005PhRvD..72d3514W}%
  \BibitemOpen
  \bibfield  {author} {\bibinfo {author} {\bibfnamefont {S.}~\bibnamefont
  {{Weinberg}}},\ }\href {\doibase 10.1103/PhysRevD.72.043514} {\bibfield
  {journal} {\bibinfo  {journal} {\prd}\ }\textbf {\bibinfo {volume} {72}},\
  \bibinfo {eid} {043514} (\bibinfo {year} {2005})}\BibitemShut {NoStop}%
\bibitem [{\citenamefont {Riemer-Sorensen}(2018)}]{bayesian}%
  \BibitemOpen
  \bibfield  {author} {\bibinfo {author} {\bibfnamefont {S.}~\bibnamefont
  {Riemer-Sorensen}},\ }\href@noop {} {\enquote {\bibinfo {title} {{LCDM and
  Beyond: Cosmology Tools in Theory and in Practice: "Statistics and model
  selection in cosmology"}},}\ }\bibinfo {howpublished}
  {\url{http://icg.port.ac.uk/~jschewts/cantata/L5/Statistics_Notes.pdf}}
  (\bibinfo {year} {2018}),\ \bibinfo {note} {accessed: 2018-05-18}\BibitemShut
  {NoStop}%
\bibitem [{\citenamefont {Rasmussen}\ and\ \citenamefont
  {Williams}(2005)}]{ML}%
  \BibitemOpen
  \bibfield  {author} {\bibinfo {author} {\bibfnamefont {C.~E.}\ \bibnamefont
  {Rasmussen}}\ and\ \bibinfo {author} {\bibfnamefont {C.~K.~I.}\ \bibnamefont
  {Williams}},\ }\href@noop {} {\emph {\bibinfo {title} {Gaussian Processes for
  Machine Learning (Adaptive Computation and Machine Learning)}}}\ (\bibinfo
  {publisher} {The MIT Press},\ \bibinfo {year} {2005})\BibitemShut {NoStop}%
\bibitem [{skl(2018)}]{sklearn}%
  \BibitemOpen
  \href@noop {} {\enquote {\bibinfo {title} {{Scikit-learn 0.19.1
  documentation: Gaussian Processes}},}\ }\bibinfo {howpublished}
  {\url{http://scikit-learn.org/stable/modules/gaussian_process.html}}
  (\bibinfo {year} {2018}),\ \bibinfo {note} {accessed: 10-06-2019}\BibitemShut
  {NoStop}%
\bibitem [{\citenamefont {Hamilton}(2000)}]{PaperFFTLOG}%
  \BibitemOpen
  \bibfield  {author} {\bibinfo {author} {\bibfnamefont {A.~J.~S.}\
  \bibnamefont {Hamilton}},\ }\href {\doibase 10.1046/j.1365-8711.2000.03071.x}
  {\bibfield  {journal} {\bibinfo  {journal} {Mon. Not. Roy. Astron. Soc.}\
  }\textbf {\bibinfo {volume} {312}},\ \bibinfo {pages} {257} (\bibinfo {year}
  {2000})},\ \Eprint {http://arxiv.org/abs/astro-ph/9905191}
  {arXiv:astro-ph/9905191 [astro-ph]} \BibitemShut {NoStop}%
%%CITATION = ASTRO-PH/9905191;%%
\bibitem [{pyt(2016)}]{pythonwrapperFFTLOG}%
  \BibitemOpen
  \href@noop {} {\enquote {\bibinfo {title} {{fftlog - python wrapper
  FFTLog}},}\ }\bibinfo {howpublished} {\url{https://github.com/prisae/fftlog}}
  (\bibinfo {year} {2016}),\ \bibinfo {note} {accessed: 01-09-2019}\BibitemShut
  {NoStop}%
\bibitem [{\citenamefont {{C{\'e}spedes}}\ \emph {et~al.}(2012)\citenamefont
  {{C{\'e}spedes}}, \citenamefont {{Atal}},\ and\ \citenamefont
  {{Palma}}}]{EFTCS2}%
  \BibitemOpen
  \bibfield  {author} {\bibinfo {author} {\bibfnamefont {S.}~\bibnamefont
  {{C{\'e}spedes}}}, \bibinfo {author} {\bibfnamefont {V.}~\bibnamefont
  {{Atal}}}, \ and\ \bibinfo {author} {\bibfnamefont {G.~A.}\ \bibnamefont
  {{Palma}}},\ }\href {\doibase 10.1088/1475-7516/2012/05/008} {\bibfield
  {journal} {\bibinfo  {journal} {Journal of Cosmology and Astroparticle
  Physics}\ }\textbf {\bibinfo {volume} {5}},\ \bibinfo {eid} {008} (\bibinfo
  {year} {2012})},\ \Eprint {http://arxiv.org/abs/1201.4848} {arXiv:1201.4848
  [hep-th]} \BibitemShut {NoStop}%
\bibitem [{\citenamefont {{Adshead}}\ and\ \citenamefont
  {{Hu}}(2014)}]{2014PhRvD..89h3531A}%
  \BibitemOpen
  \bibfield  {author} {\bibinfo {author} {\bibfnamefont {P.}~\bibnamefont
  {{Adshead}}}\ and\ \bibinfo {author} {\bibfnamefont {W.}~\bibnamefont
  {{Hu}}},\ }\href {\doibase 10.1103/PhysRevD.89.083531} {\bibfield  {journal}
  {\bibinfo  {journal} {\prd}\ }\textbf {\bibinfo {volume} {89}},\ \bibinfo
  {eid} {083531} (\bibinfo {year} {2014})},\ \Eprint
  {http://arxiv.org/abs/1402.1677} {arXiv:1402.1677 [astro-ph.CO]} \BibitemShut
  {NoStop}%
\bibitem [{\citenamefont {{Cannone}}\ \emph {et~al.}(2014)\citenamefont
  {{Cannone}}, \citenamefont {{Bartolo}},\ and\ \citenamefont
  {{Matarrese}}}]{2014PhRvD..89l7301C}%
  \BibitemOpen
  \bibfield  {author} {\bibinfo {author} {\bibfnamefont {D.}~\bibnamefont
  {{Cannone}}}, \bibinfo {author} {\bibfnamefont {N.}~\bibnamefont
  {{Bartolo}}}, \ and\ \bibinfo {author} {\bibfnamefont {S.}~\bibnamefont
  {{Matarrese}}},\ }\href {\doibase 10.1103/PhysRevD.89.127301} {\bibfield
  {journal} {\bibinfo  {journal} {\prd}\ }\textbf {\bibinfo {volume} {89}},\
  \bibinfo {eid} {127301} (\bibinfo {year} {2014})},\ \Eprint
  {http://arxiv.org/abs/1402.2258} {arXiv:1402.2258 [astro-ph.CO]} \BibitemShut
  {NoStop}%
\bibitem [{\citenamefont {{Handley}}\ \emph
  {et~al.}(2015{\natexlab{a}})\citenamefont {{Handley}}, \citenamefont
  {{Hobson}},\ and\ \citenamefont {{Lasenby}}}]{Polychord}%
  \BibitemOpen
  \bibfield  {author} {\bibinfo {author} {\bibfnamefont {W.~J.}\ \bibnamefont
  {{Handley}}}, \bibinfo {author} {\bibfnamefont {M.~P.}\ \bibnamefont
  {{Hobson}}}, \ and\ \bibinfo {author} {\bibfnamefont {A.~N.}\ \bibnamefont
  {{Lasenby}}},\ }\href {\doibase 10.1093/mnrasl/slv047} {\bibfield  {journal}
  {\bibinfo  {journal} {Mon. Not. Roy. Astron. Soc.}\ }\textbf {\bibinfo
  {volume} {450}},\ \bibinfo {pages} {L61} (\bibinfo {year}
  {2015}{\natexlab{a}})},\ \Eprint {http://arxiv.org/abs/1502.01856}
  {arXiv:1502.01856} \BibitemShut {NoStop}%
\bibitem [{\citenamefont {{Handley}}\ \emph
  {et~al.}(2015{\natexlab{b}})\citenamefont {{Handley}}, \citenamefont
  {{Hobson}},\ and\ \citenamefont {{Lasenby}}}]{Polychord2}%
  \BibitemOpen
  \bibfield  {author} {\bibinfo {author} {\bibfnamefont {W.~J.}\ \bibnamefont
  {{Handley}}}, \bibinfo {author} {\bibfnamefont {M.~P.}\ \bibnamefont
  {{Hobson}}}, \ and\ \bibinfo {author} {\bibfnamefont {A.~N.}\ \bibnamefont
  {{Lasenby}}},\ }\href {\doibase 10.1093/mnras/stv1911} {\bibfield  {journal}
  {\bibinfo  {journal} {Mon. Not. Roy. Astron. Soc.}\ }\textbf {\bibinfo
  {volume} {453}},\ \bibinfo {pages} {4384} (\bibinfo {year}
  {2015}{\natexlab{b}})},\ \Eprint {http://arxiv.org/abs/1506.00171}
  {arXiv:1506.00171 [astro-ph.IM]} \BibitemShut {NoStop}%
\bibitem [{\citenamefont {Lewis}(2019)}]{getdist}%
  \BibitemOpen
  \bibfield  {author} {\bibinfo {author} {\bibfnamefont {A.}~\bibnamefont
  {Lewis}},\ }\href@noop {} {\  (\bibinfo {year} {2019})},\ \Eprint
  {http://arxiv.org/abs/1910.13970} {arXiv:1910.13970 [astro-ph.IM]}
  \BibitemShut {NoStop}%
%%CITATION = ARXIV:1910.13970;%%
\bibitem [{\citenamefont {Lewis}(2013)}]{Lewis:2013hha}%
  \BibitemOpen
  \bibfield  {author} {\bibinfo {author} {\bibfnamefont {A.}~\bibnamefont
  {Lewis}},\ }\href {\doibase 10.1103/PhysRevD.87.103529} {\bibfield  {journal}
  {\bibinfo  {journal} {Phys. Rev.}\ }\textbf {\bibinfo {volume} {D87}},\
  \bibinfo {pages} {103529} (\bibinfo {year} {2013})},\ \Eprint
  {http://arxiv.org/abs/1304.4473} {arXiv:1304.4473 [astro-ph.CO]} \BibitemShut
  {NoStop}%
%%CITATION = ARXIV:1304.4473;%%
\bibitem [{\citenamefont {{Torrado}}\ and\ \citenamefont
  {{Lewis}}(2020)}]{2020arXiv200505290T}%
  \BibitemOpen
  \bibfield  {author} {\bibinfo {author} {\bibfnamefont {J.}~\bibnamefont
  {{Torrado}}}\ and\ \bibinfo {author} {\bibfnamefont {A.}~\bibnamefont
  {{Lewis}}},\ }\href@noop {} {\bibfield  {journal} {\bibinfo  {journal} {arXiv
  e-prints}\ ,\ \bibinfo {eid} {arXiv:2005.05290}} (\bibinfo {year} {2020})},\
  \Eprint {http://arxiv.org/abs/2005.05290} {arXiv:2005.05290 [astro-ph.IM]}
  \BibitemShut {NoStop}%
\bibitem [{\citenamefont {{Cartis}}\ \emph
  {et~al.}(2018{\natexlab{a}})\citenamefont {{Cartis}}, \citenamefont
  {{Fiala}}, \citenamefont {{Marteau}},\ and\ \citenamefont
  {{Roberts}}}]{2018arXiv180400154C}%
  \BibitemOpen
  \bibfield  {author} {\bibinfo {author} {\bibfnamefont {C.}~\bibnamefont
  {{Cartis}}}, \bibinfo {author} {\bibfnamefont {J.}~\bibnamefont {{Fiala}}},
  \bibinfo {author} {\bibfnamefont {B.}~\bibnamefont {{Marteau}}}, \ and\
  \bibinfo {author} {\bibfnamefont {L.}~\bibnamefont {{Roberts}}},\ }\href@noop
  {} {\bibfield  {journal} {\bibinfo  {journal} {arXiv e-prints}\ ,\ \bibinfo
  {eid} {arXiv:1804.00154}} (\bibinfo {year} {2018}{\natexlab{a}})},\ \Eprint
  {http://arxiv.org/abs/1804.00154} {arXiv:1804.00154 [math.OC]} \BibitemShut
  {NoStop}%
\bibitem [{\citenamefont {{Cartis}}\ \emph
  {et~al.}(2018{\natexlab{b}})\citenamefont {{Cartis}}, \citenamefont
  {{Roberts}},\ and\ \citenamefont {{Sheridan-Methven}}}]{2018arXiv181211343C}%
  \BibitemOpen
  \bibfield  {author} {\bibinfo {author} {\bibfnamefont {C.}~\bibnamefont
  {{Cartis}}}, \bibinfo {author} {\bibfnamefont {L.}~\bibnamefont {{Roberts}}},
  \ and\ \bibinfo {author} {\bibfnamefont {O.}~\bibnamefont
  {{Sheridan-Methven}}},\ }\href@noop {} {\bibfield  {journal} {\bibinfo
  {journal} {arXiv e-prints}\ ,\ \bibinfo {eid} {arXiv:1812.11343}} (\bibinfo
  {year} {2018}{\natexlab{b}})},\ \Eprint {http://arxiv.org/abs/1812.11343}
  {arXiv:1812.11343 [math.OC]} \BibitemShut {NoStop}%
\bibitem [{\citenamefont {J.~D.~Powell}(2009)}]{BOBYQA}%
  \BibitemOpen
  \bibfield  {author} {\bibinfo {author} {\bibfnamefont {M.}~\bibnamefont
  {J.~D.~Powell}},\ }\href@noop {} {\bibfield  {journal} {\bibinfo  {journal}
  {Technical Report, Department of Applied Mathematics and Theoretical
  Physics}\ } (\bibinfo {year} {2009})}\BibitemShut {NoStop}%
\bibitem [{\citenamefont {{Akaike}}(1974)}]{1100705}%
  \BibitemOpen
  \bibfield  {author} {\bibinfo {author} {\bibfnamefont {H.}~\bibnamefont
  {{Akaike}}},\ }\href {\doibase 10.1109/TAC.1974.1100705} {\bibfield
  {journal} {\bibinfo  {journal} {IEEE Transactions on Automatic Control}\
  }\textbf {\bibinfo {volume} {19}},\ \bibinfo {pages} {716} (\bibinfo {year}
  {1974})}\BibitemShut {NoStop}%
\bibitem [{\citenamefont {{Handley}}\ \emph
  {et~al.}(2015{\natexlab{c}})\citenamefont {{Handley}}, \citenamefont
  {{Hobson}},\ and\ \citenamefont {{Lasenby}}}]{2015MNRAS.453.4384H}%
  \BibitemOpen
  \bibfield  {author} {\bibinfo {author} {\bibfnamefont {W.~J.}\ \bibnamefont
  {{Handley}}}, \bibinfo {author} {\bibfnamefont {M.~P.}\ \bibnamefont
  {{Hobson}}}, \ and\ \bibinfo {author} {\bibfnamefont {A.~N.}\ \bibnamefont
  {{Lasenby}}},\ }\href {\doibase 10.1093/mnras/stv1911} {\bibfield  {journal}
  {\bibinfo  {journal} {Mon. Not. Roy. Astron. Soc.}\ }\textbf {\bibinfo
  {volume} {453}},\ \bibinfo {pages} {4384} (\bibinfo {year}
  {2015}{\natexlab{c}})},\ \Eprint {http://arxiv.org/abs/1506.00171}
  {arXiv:1506.00171 [astro-ph.IM]} \BibitemShut {NoStop}%
\bibitem [{\citenamefont {{Hu}}\ and\ \citenamefont
  {{Torrado}}(2015)}]{2015PhRvD..91f4039H}%
  \BibitemOpen
  \bibfield  {author} {\bibinfo {author} {\bibfnamefont {B.}~\bibnamefont
  {{Hu}}}\ and\ \bibinfo {author} {\bibfnamefont {J.}~\bibnamefont
  {{Torrado}}},\ }\href {\doibase 10.1103/PhysRevD.91.064039} {\bibfield
  {journal} {\bibinfo  {journal} {\prd}\ }\textbf {\bibinfo {volume} {91}},\
  \bibinfo {eid} {064039} (\bibinfo {year} {2015})},\ \Eprint
  {http://arxiv.org/abs/1410.4804} {arXiv:1410.4804} \BibitemShut {NoStop}%
\end{thebibliography}%

\end{document}